\documentclass[11pt]{article}
\usepackage{amssymb}
\usepackage{latexsym}
\usepackage{amsmath}
\usepackage{graphicx}
\usepackage{listings}

\def\al{\alpha}
\def\L{\Lambda}
\def\l{\lambda}

\title{\LARGE{Static spherically symmetric Einstein-Yang-Mills-dilaton black hole and its thermodynamics }}
\author{M. M. Stetsko\footnote{E-mail: mstetsko@gmail.com}\
\\
  {\small Department for Theoretical Physics, Ivan Franko National University of Lviv,}\\
{\small 12 Drahomanov Str., Lviv, UA-79005, Ukraine
         }}
\begin{document}
\maketitle

{\abstract{A static black hole with spherical symmetry is obtained and examined in the framework of Einstein-Yang-Mills-dilaton theory. The obtained black hole solution allowed us to derive and investigate entropy, temperature and heat capacity. To better examine the thermodynamics of the black hole extended phase space is also used. On this ground the equation of state is obtained and studied. We have also investigated the Gibbs free energy and it is shown that below the critical temperature the system demonstrates phase transitions of the first as well as of the zeroth order which is notable feature for other types of dilaton black holes. At the end critical exponents for the black hole are calculated.}}

\section{Introduction}
Black holes are one of the most fascinating objects of investigations in General Relativity as well as in other approaches to Gravity \cite{Capo_PRep11,Clifton_PhysRept2012,Heisenberg}. This deep and longstanding interest to various aspects of Black Holes' Physics can be explained not only by their great importance for Astrophysics \cite{Will_LRR2014}, but also due to numerous applications of the techniques and methods initially developed in Gravity to various non gravitational systems. For instance, the application of these methods is motivated by AdS/CFT correspondence \cite{Maldacena_ATMP98,Witten_ATMP98}, which established the correspondence between Gravity in the bulk and Conformal Field Theory on its boundary.

Apart of well known vacuum solutions such as Schwarzschild or Kerr black holes and which can be treated as one of the simplest toy models, the great interest is attracted to the black holes where additional material fields are taken into consideration. The most widely investigated are scalar fields, which appear for instance in Scalar-Tensor Theories of Gravity \cite{Capo_PRep11,Heisenberg} or have mainly String Theory origin as well-known dilaton or axion fields. Deep interest is attracted also to black holes  with electromagnetic field, for which the simplest solution, namely the Reissner-Nordstrom black hole is known from the early days of General Relativity. The black holes solutions with both scalar and Maxwell fields were also considered, in particular in the framework of Einstein-Maxwell-dilaton theory \cite{Gibbons_NPB88,Garfinkle_PRD91,Witten_PRD91,Gregory_PRD93,Rakhmanov_PRD94,
Poletti_PRD94,Chan_NPB95,Cai_PRD96,Gao_PRD04,Yazadjiev_CQG05,Astefanesei_PRD06,
Mann_JHEP06,Kunz_PLB06,Brihaye_CQG07_1x,Charmousis_PRD09, Sheykhi_PRD07,Sheykhi_PLB08,Sheykhi_PRD14,Kord_PRD15,
Hendi_PRD15,Dehyadegari_PRD17,Stetsko_EPJC19}.

Nonabelian gauge fields can be treated as a some kind of generalization  of the Maxwell one and thus the investigation of the black holes' solutions with nonabelian fields is a task of great importance. We also point out here that black holes with Yang-Mills fields has also relatively long history, but due to complicated structure of nonabelian fields they were not so widely studied in comparison with the Maxwell field. The first black hole's solution with nonabelian field with $SO(3)$ gauge group was obtained and examined by Yasskin in 1975 \cite{Yasskin_PRD75}, here we also point out the work of Kasuya where nonableian rotating dyon black hole solution was derived \cite{Kasuya_PRD82}. Several years later soliton \cite{Bartnik_PRL88} and black hole's \cite{Bizon_PRL90,Volkov_JETP89} solutions were obtained in Einstein-Yang-Mills theory with $SU(2)$ gauge group. The latter solutions were observed to be unstable \cite{Straumann_PLB90}, but they stimulated the deep interest to solutions in Einstein-Yang-Mills theory \cite{Torii_PRD95,Volkov_PRD96, Volkov_PhRp98,Mavromatos_JMP98}. Stable black holes were shown to exist in case of anti-de Sitter space \cite{Winstanley_CQG99,Bjoraker_PRL00,vdBij_PLB02}. The existence of stable solutions in Einstein-Yang-Mills theory stimulated deeper interest for seeking of new solutions and their examination. Namely, Wu-Yang ansatz was applied in higher dimensional case and so called magnetically charged solutions were derived \cite{Mazhari_PRD07,Mazhari_JCAP08}. Black holes in Einsteinian gravity with nonlinear Yang-Mills field of Born-Infeld type \cite{Mazhari_PRD08} as well as their extension to dilaton theory \cite{Mazhari_GRG10} were studied. Wu-Yang ansatz was also utilized in case of third order Lovelock gravity with nonabelian gauge field \cite{Mazhari_PLB08_2}. Black holes in $F(R)$ gravity coupled with Yang-Mills field as well as in  the presence of a scalar field were obtained \cite{Mazhari_PRD11,Mazhari_EPJC13}. Topological black holes in the framework of Einstein-Yang-Mills theory were derived and investigated \cite{Bostani_MPLA10,Dehghani_IJMPD10}. Black hole's solution in Einstein-dilaton theory with $SU(2)$ nonabelian field was obtained in \cite{Radu_CQG05_1}. Coloured black holes with higher order curvature terms were also derived and studied \cite{Brihaye_PLB03}. Higher dimensional Einstein-Yang-Mills black hole with gauge field terms belonging to higher order Yang-Mills hierarchy were studied \cite{Brihaye_PRD07}. Nonabelian black hole with superconducting  horizons in AdS space was examined \cite{Manvelyan_PLB09}. Soliton and black hole solutions of ${\mathfrak{su}}(N)$ Yang-Mills black hole in AdS space were derived \cite{Baxter_PRD07}. Particle-like and black hole solutions were considered in Einstein-Yang-Mills-Chern-Simons theory \cite{Lerida_PRD09}. Five dimensional radiating black holes were studied in case of Einstein-Yang-Mills-Gauss-Bonnet gravity \cite{Ghosh_PLB11} and rotating solution was derived \cite{Ghosh_EPJC14}. Five dimensional nonabelian black holes in the framowork of supergravity was obtained and examined \cite{Mann_PRD06,Cvetic_PRD10}. Nonabelian solutions with NUT charges were obtained and investigated \cite{Radu_PRD03}. Four- and five-dimensional charged black holes were studied in colored Lifshitz space-time \cite{Fan_PLB15}. Magnetic black holes in nonminimal Einstein-Yang-Mills theory were considered \cite{Balakin_PRD16}. Charged Einstein-Yang-Mills black hole in the framework of gravity's rainbow was examined and its thermodynamics was studied \cite{Hendi_PLB18}. Black hole's solution with nonlinear abelian and nonabelian fields was obtained within the massive gravity, its thermodynamics and quasi-normal modes were studied \cite{Hendi_JHEP19}. Black hole with Yang-Mills field was examined in dimensionally continued gravity \cite{Ali_PRD19}.

In our work we consider a static black hole with spherical symmetry in Einstein-Yang-Mills-dilaton theory. Black holes with nonabelian and dilaton fields were studied in various papers, in particular  \cite{Mazhari_GRG10,Radu_CQG05_1,Mann_PRD06,
Kanti_PRD00,Kleihaus_PRD04,Aschbacher_PRD03,Aprile_JHEP11,
Feng_PLB15,Kleihaus_CQG16} and some others. The character of solution one is to obtain in case of Yang-Mills theory substantially depends on the chosen gauge group. In this paper we take $SO(n)$ group and the nonabelian field potential is chosen in the so-called magnetic Wu-Yang form \cite{Mazhari_PRD07,Mazhari_JCAP08}. We point out that $SO(n)$ nonabelian dilaton black hole's solution was obtained in \cite{Mazhari_GRG10}, but that solution is of the so-called Bertotti-Robinson (BR) class, which in case of a charged black hole is closer to the extremal case rather than to the nonextreme RN solution. In our work the solution we obtain is not of the BR type, it is closer to nonextreme solution. To obtain an exact solution we also choose the dilation potential in a Liouville-type form. It should be pointed out here that as far as we convinced in most of the papers were dilaton black holes' solutions are obtained, especially with additional material fields,  potentials of the Liouville-type were used. In our case the chosen form of the Liouville potential has some peculiarities, namely one of its terms is chosen to take into account cosmological constant and the other one will be related to the gauge filed. In contrast, in the paper \cite{Mazhari_GRG10} the dilaton potential is not introduced, because the solution as it has just been noted is of the Bertotti-Robinson class and due to the fact that the cosmological constant is not taken into account. We also point out that black holes with $SO(n)$ gauge field were examined also in \cite{Mazhari_PRD08,Mazhari_PLB08_2,Mazhari_PRD11}, but in those works the gravitational part of the action was different and the dilaton field was not taken into account, although the gauge potential was the same as in \cite{Mazhari_PRD07,Mazhari_GRG10}. From the other side the solutions which are close to the mentioned ones were obtained in \cite{Dehghani_IJMPD10},where topological solutions were derived.  Using the obtained solution we also consider thermodynamics of the black hole, namely we obtain and investigate temperature and heat capacity and the latter one, as it is known, allows us to study thermal stability of the black hole. Then we use extended thermodynamics concept \cite{Kastor_CQG09,Kubiznak_JHEP12} and obtain thermal equation of state and investigate critical behaviour. We also note here that the extended thermodynamics approach which is widely applied to the black holes with the Maxwell field is not so actively used in case of the black holes with the Yang-Mills fields, this work is one of those where the extended thermodynamics of the black hole is extensively studied.

This paper is organized as follows. In the next section we obtain the equations of motion for Einstein-Yang-Mills-dilaton theory, then we derive static black hole's solution and analyze it.  The third section is devoted to the standard black hole thermodynamics, namely, we calculate temperature, entropy and write the first law of black hole thermodynamics, we also obtain and examine heat capacity of the black hole. In the forth section we use the extended thermodynamics concept, namely we derive and investigate the equation of state and we also study the behaviour of the Gibbs free energy. In the fifth section we calculate critical exponents. The sixth sections contains some conclusions.

\section{Field equations for Einstein-dilaton-Yang-Mills theory and static black hole solution}
We consider $n+1$--dimensional ($n\geqslant 3$) Einstein-dilaton-Yang-Mills theory where dilaton and Yang-Mills fields are coupled and coupling for these fields to gravity is minimal. The action integral for the system takes the form:
\begin{eqnarray}\label{action_int}
 S=\frac{1}{16\pi}\int_{{\cal M}} {\rm d}^{n+1}x\sqrt{-g}\left(R-\frac{4}{n-1}\nabla^{\mu}\Phi\nabla_{\mu}\Phi-V(\Phi)-e^{-4\alpha\Phi/(n-1)}Tr(F^{(a)}_{\mu\nu}F^{(a)\mu\nu})\right)+\frac{1}{8\pi}\int_{\partial{\cal M}} d^{n}x\sqrt{-h}K
\end{eqnarray}
and here $g$ denotes the determinant of the metric tensor $g_{\mu\nu}$, $R$ is the scalar curvature, $\Phi$ is the dilaton field, $V(\Phi)$ denotes the dilaton potential, $F^{(a)}_{\mu\nu}$ is the gauge field tensor, parameter $\al$ is the coupling constant between dilaton and Yang-Mills fields. The second integral in the action (\ref{action_int}) is the boundary Gibbons-Hawking-York (GHY) term which is introduced to make the variation problem well-defined. In the GHY-term $h$ denotes the determinant of the boundary metric $h_{\mu\nu}$ and $K$ is the trace of extrinsic curvature tensor. 

The gauge field tensor is defined in the following way:
\begin{equation}\label{gauge_field}
F^{(a)}_{\mu\nu}=\partial_{\mu}A^{(a)}_{\nu}-\partial_{\nu}A^{(a)}_{\mu}+\frac{1}{2\bar{\kappa}}C^{(a)}_{(b)(c)}A^{(b)}_{\mu}A^{(c)}_{\nu},
\end{equation} 
where $A^{(a)}$ is the gauge field potential, $C^{(a)}_{(b)(c)}$ are the structure constants for the corresponding gauge group and $\bar{\kappa}$ is the coupling constant for the nonabelian field. In this paper we consider the gauge group to be $SO(n)$.

Having used the least action principle we derive equations of motion for the system described by the action (\ref{action_int}) which can be written in the form:
\begin{equation}\label{einstein}
R_{\mu\nu}=\frac{g_{\mu\nu}}{n-1}\left(V(\Phi)-e^{-4\al\Phi/(n-1)}Tr(F^{(a)}_{\rho\sigma}F^{(a)\rho\sigma})\right)+\\
\frac{4}{n-1}\partial_{\mu}\Phi\partial_{\nu}\Phi+2e^{-4\al\Phi/(n-1)}
Tr(F^{(a)}_{\mu\sigma}{F^{(a)\sigma}_{\nu}});
\end{equation}
\begin{eqnarray}
\nabla_{\mu}\nabla^{\mu}\Phi=\frac{n-1}{8}\frac{\partial V}{\partial \Phi}-\frac{\al}{2}e^{-4\al\Phi/(n-1)}Tr(F^{(a)}_{\rho\sigma}F^{(a)\rho\sigma})\label{scal_eq};\\
\nabla_{\mu}(e^{-4\al\Phi/(n-1)}F^{(a)\mu\nu})+\frac{1}{\bar{\kappa}}e^{-4\al\Phi/(n-1)}C^{(a)}_{(b)(c)}A^{(b)}_{\mu}F^{(c)\mu\nu}=0\label{YM_eq}.
\end{eqnarray}
In this work we consider a static spherically symmetric solution of the field equations, so we take the metric in the following form:
\begin{equation}\label{metric}
 ds^2=-W(r)dt^2+\frac{dr^2}{W(r)}+r^2R^2(r)d\Omega^2_{n-1},
\end{equation}
where $d\Omega^2_{n-1}$ is the line element of $n-1$-dimensional unit hypersphere.

For the gauge field potential $A^{(a)}_{\mu}$ we choose Wu-Yang ansatz and write components of the gauge potential form:
\begin{equation}\label{gauge_pot}
{\bf A}^{(a)}=\frac{q}{r^2}C^{(a)}_{(i)(j)}x^{i}dx^{j}, \quad r^2=\sum^{n}_{j=1}x^2_j,
\end{equation}
and here indices $a$, $i$ and $j$ run the following ranges $2\leqslant j+1<i\leqslant n$ and $1\leqslant a\leqslant n(n-1)/2$ and here new parameter $q$ is taken to be equal to the Yang-Mills coupling constant $\bar{\kappa}$ ($q=\bar{\kappa}$). The coordinates $x_i$ in the relation (\ref{gauge_pot}) can be represented by virtue of the following relation:
\begin{eqnarray}
\nonumber x_1=r\cos{\chi_{n-1}}\sin{\chi_{n-2}}\ldots\sin{\chi_1},\quad x_2=r\sin{\chi_{n-1}}\sin{\chi_{n-2}}\ldots\sin{\chi_1},\\
\nonumber  x_3=r\cos{\chi_{n-2}}\sin{\chi_{n-3}}\ldots\sin{\chi_1},\quad x_4=r\sin{\chi_{n-2}}\sin{\chi_{n-3}}\ldots\cos{\chi_1},\\
\nonumber \cdots \quad\quad\\
x_n=r\cos{\chi_1}
\end{eqnarray}
The angular variabels $\chi_1,\ldots,\chi_{n-1}$ also allow to define the line element of the unit hypersphere in metric (\ref{metric}), namely we write:
\begin{equation}
d\Omega^2_{n-1}=d\chi^2_{1}+\sum^{n-1}_{j=2}\prod^{j-1}_{i=1}\sin^2{\chi_{i}}d\chi^2_{j},
\end{equation}
and the angular variables vary in the ranges $0\leqslant\chi_{i}\leqslant\pi, i=1,\ldots,n-2$, $0\leqslant\chi_{n-1}\leqslant 2\pi$.

Having used the Wu-Yang ansatz for the gauge potential (\ref{gauge_pot}) and definition of the gauge field (\ref{gauge_field}) one can verify that the gauge field equations (\ref{YM_eq}) are satisfied. One can calculate gauge field invariant which takes the form:
\begin{equation}
Tr(F^{(a)}_{\rho\sigma}F^{(a)\rho\sigma})=(n-1)(n-2)\frac{q^2}{r^4R^4}.
\end{equation}
To solve field equations (\ref{einstein}) and (\ref{scal_eq}) an evident form of the dilaton potential $V(\Phi)$ should be given. We choose the potential $V(\Phi)$ in the so-called Liouville form:
\begin{equation}
V(\Phi)=\Lambda e^{\lambda\Phi}+\Lambda_1 e^{\lambda_1\Phi}+\Lambda_2e^{\lambda_2\Phi},
\end{equation} 
where parameters $\l$ $\l _i$ and $\Lambda_i$, $i=1,2$ will be obtained to fulfill the equations of motion. We point out here that Liouville-type potentials were used in numerous papers where dilaton field was coupled to Maxwell one \cite{Sheykhi_PRD07,Stetsko_EPJC19}. Here we also use the following ansatz for the function $R(r)$:
\begin{equation}\label{ansatz_R}
R(r)=e^{2\al\Phi/(n-1)}.
\end{equation}
Having used this ansatz we obtain the solution of the field equations in the following form:
\begin{eqnarray}
\nonumber W(r)=-mr^{1+(1-n)(1-\gamma)}+\frac{(n-2)(1+\al^2)^2}{(1-\al^2)(\al^2+n-2)}b^{-2\gamma}r^{2\gamma}-\\\frac{\Lambda(1+\al^2)^2}{(n-1)(n-\al^2)}b^{2\gamma}r^{2(1-\gamma)}+\frac{(n-2)q^2(1+\al^2)^2}{(\al^2-1)(n+3\al^2-4)}b^{-6\gamma}r^{6\gamma-2}\label{metric_W}
\end{eqnarray}
where $\gamma=\al^2/(1+\al^2)$. And the dilaton field $\Phi$ takes the form:
\begin{equation}
\Phi(r)=\frac{\al(n-1)}{2(1+\al^2)}\ln{\left(\frac{b}{r}\right)}.
\end{equation}
Here we note that in the written above expression for the metric function (\ref{metric_W}) the parameter $m$ is an integration constant related to black hole's mass, and the other parameter $b$ is also an integration constant, but its physical meaning is not so transparent, from its form we suppose that it might be related to some rescaling properties for the obtained solution.  
As it has been noted above the parameters of Liouville potential are taken to satisfy the equations of motion. It can be easily verified that these parameters should take the form:
\begin{eqnarray}
\lambda_0=\frac{4\al}{(n-1)},\quad \lambda_1=\frac{4}{\al(n-1)}, \quad \lambda_2=\frac{4(2-\al^2)}{\al(n-1)}\\
\Lambda_1=\frac{\alpha^2(n-1)(n-2)}{b^2(\al^2-1)}, \quad \Lambda_2=-\frac{\al(n-1)(n-2)}{\al^2-1}q^2b^{-4}.
\end{eqnarray}   
and the parameter $\L$ can be taken arbitrary and it will be treated as an effective cosmological constant, similarly as it was done for Einstein-Maxwell-dilaton black holes \cite{Stetsko_EPJC19}. It should be also pointed out that the metric function (\ref{metric_W}) is not well-defined when $\al=1$ this is so-called string singularity and when $\al=\sqrt{n}$. Similar peculiarities take place for the already mentioned Einstein-Maxwell-dilaton black hole  \cite{Sheykhi_PRD07,Stetsko_EPJC19} and they are related to the dilaton part. In the limit when $\alpha=0$ we arrive at the metric of Einstein-Yang-Mills solution \cite{Bostani_MPLA10}:
\begin{equation}
W(r)=1-\frac{m}{r^{n-2}}+\frac{\L}{n(n-1)}r^2-\frac{(n-2)q^2}{(n-4)r^2}.
\end{equation} 
We remark that here $n\neq 4$ and if $n=4$ the last term in the latter relation becomes logarithmic instead of the power-law one. 

\begin{figure}
\centerline{\includegraphics[scale=0.3,clip]{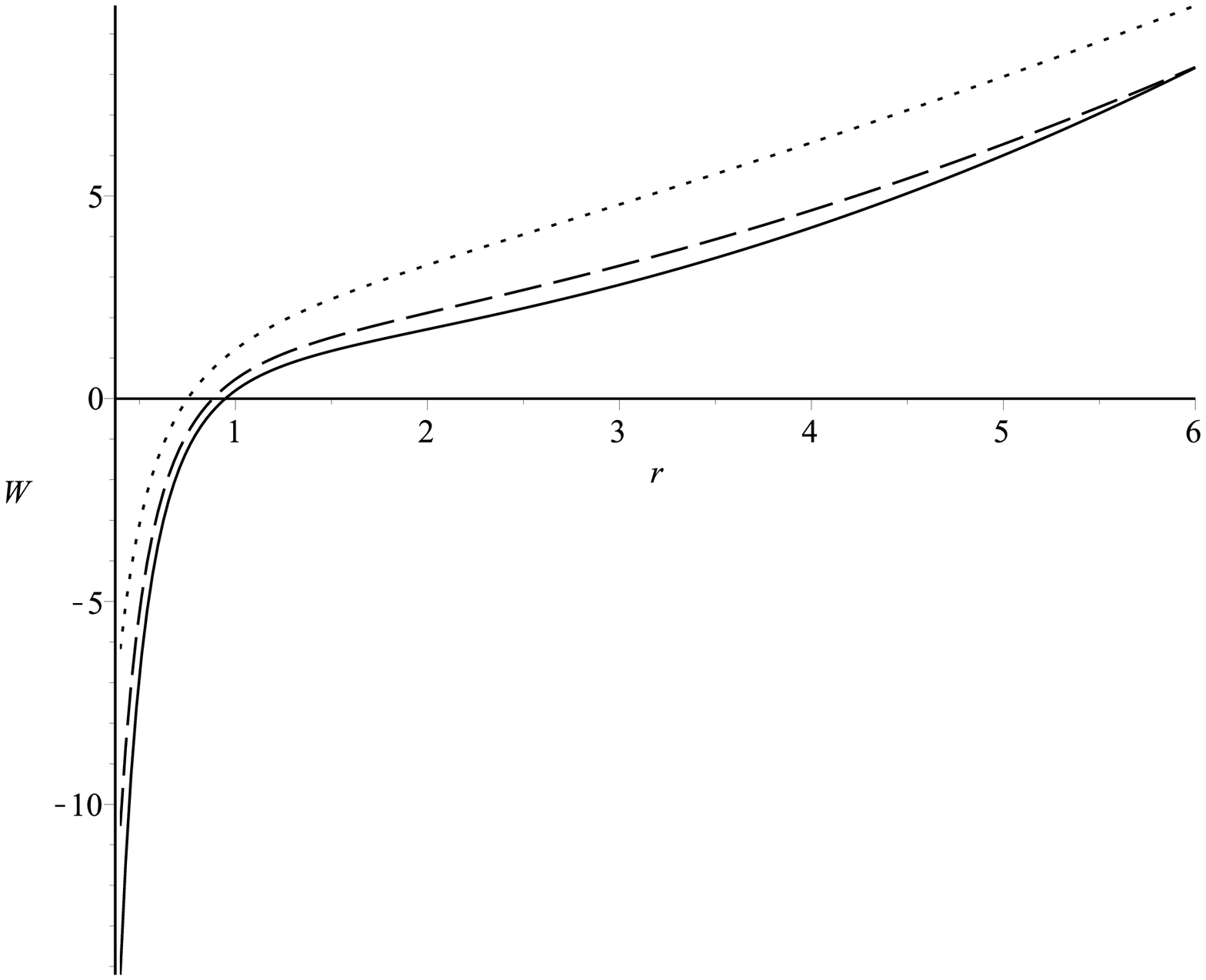}\includegraphics[scale=0.3,clip]{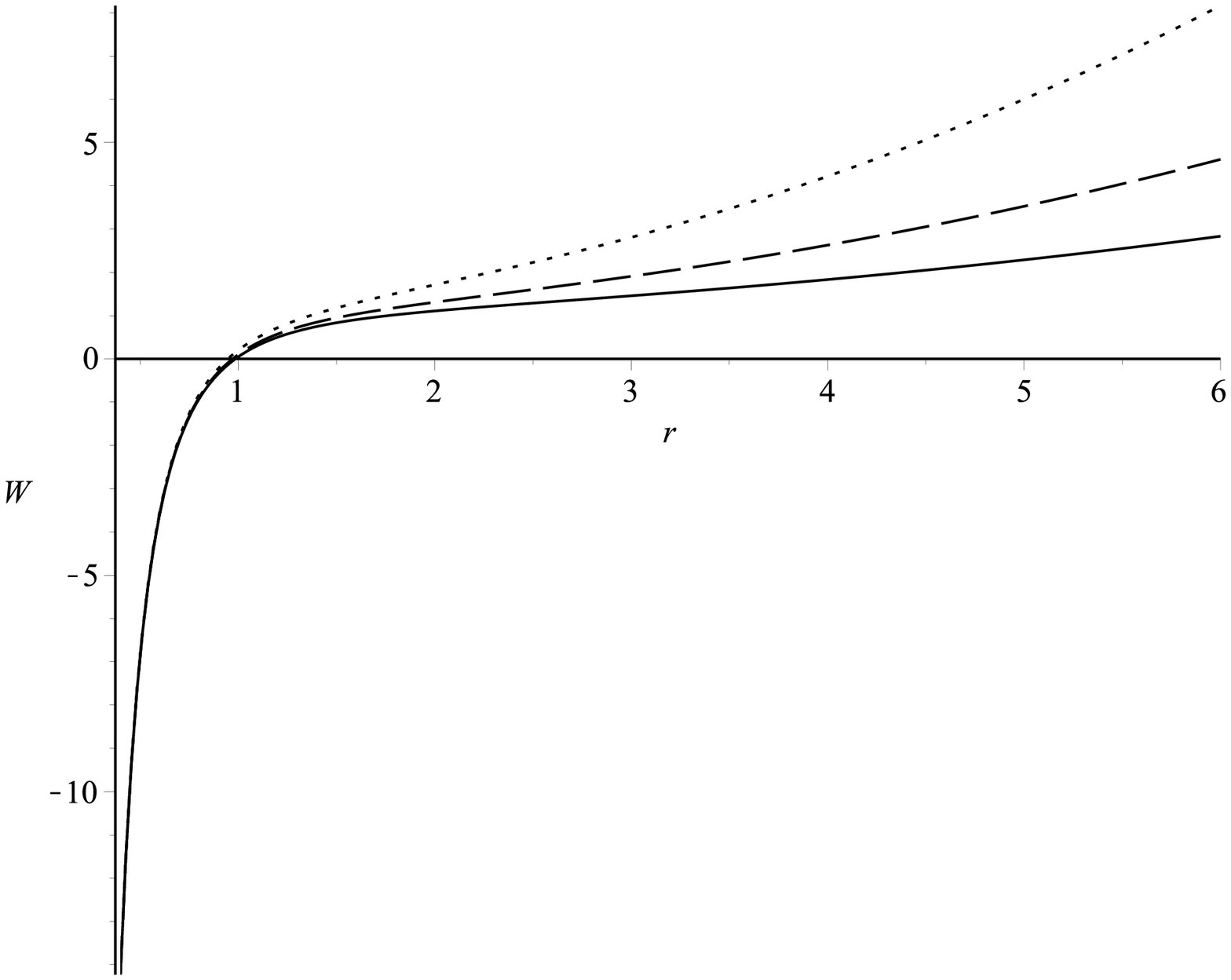}}
\caption{Metric function W(r) for different values of parameter $\al$ (the left graph) and cosmological constant $\Lambda$ (the right one). For both graphs we have taken: $n=5,b=1,m=1,q=0.2$. For the left plot the parameter $\al$ is chosen to be: $\al=0.1$, $\al=0.3$, $\al=0.5$ for solid, dashed and dotted curves respectively and $\L=-4$. For the right graph the varied parameter $\L$ is taken to be $\L=-1$, $\L=-2$ and $\L=-4$ for solid, dashed and dotted lines correspondingly, while $\al=0.1$.}\label{metric_graph}
\end{figure}

The evident form of the metric function $W(r)$ (\ref{metric_W}) is not so simple to characterize its behaviour in full generality, but nevertheless some general conclusions can be made. Firstly, when $n>3$ its behaviour for small distances ($r\rightarrow 0$)  is defined mainly by the term $\sim mr^{1+(1-n)(1-\gamma)}$ (Schwarzschild term). If $n=3$ for small values of the coupling parameter $\al$ the gauge field term in the solution (\ref{metric_W}) also might give considerable contribution, so the resulting behaviour of the metric would be defined by the relation between these two terms. For large distances (asymptotic infinity) dominating terms might be the second or the third terms in the metric function $W(r)$ depending on the value or $\al$, actually this behaviour is not exactly of the dS or AdS-type, but the metric also is not an asymptotically flat. The metric would be close to the dS or AdS type in the limit case, when the coupling parameter $\al$ is very small ($\al\rightarrow 0$), then the third term in (\ref{metric_W}) approaches the dS or AdS-type term depending on the sign of $\L$. To better understand the behaviour of the metric for some intermediate distances we show it graphically. Namely, Figure [\ref{metric_graph}] demonstrates that that  the variation of the parameter $\al$ considerably affects on behaviour of the metric function $W(r)$ for all distances and it is well explained by the fact that the parameter $\al$ is present in all the terms in the metric function $W(r)$. Whereas the influence of the parameter $\L$ becomes important for large distances, namely at the asymptotic infinity, at the same time its influence for small distances might be negligibly small, since the parameter $\L$ is present in the third term of the function $W(r)$ only and it would be substantial just for large distances.   

Coordinate and physical singularities of the metric are revealed if one examines behaviour of Kretschmann scalar:
\begin{equation}\label{Kr_scal}
R_{\mu\nu\rho\sigma}R^{\mu\nu\rho\sigma}=\left(W''\right)^2+\frac{(n-1)}{(rR)^2}\left[\left((rR)'\right)^2\left(W'\right)^2+\left(2(rR)''W-(rR)'W'\right)^2\right]+\frac{2(n-1)(n-2)}{(rR)^4}\left(1-\left((rR)'\right)^2W\right)^2.
\end{equation} 
It is easy to check that at the horizon the Kretschmann scalar (\ref{Kr_scal}) is not singular and it means that the horizon is the point of a coordinate singularity as it should be for a black hole. Taking the leading terms of the metric function $W(r)$ (\ref{metric_W}) at the origin and calculating the Kretschmann scalar we arrive at the following expression:
\begin{equation}
R_{\mu\nu\rho\sigma}R^{\mu\nu\rho\sigma}\simeq\frac{(n-1)}{(1+\al^2)^4}\left((n+1)(2+\al^2-n)^2+2(n-2)+4\al^2(2-n+2\al^2)\right)m^2r^{2(1-n)(1-\gamma)-2}.
\end{equation}
The latter expression shows singular behaviour when $r\rightarrow 0$ and as a consequence the origin ($r=0$) is the point of true physical singularity. It should also be noted that if $\al=0$ the latter relation is reduced to corresponding expression of the Kretschmann scalar near the origin for $n+1$-dimensional Schwarzschild (or Schwarzschild-Tangherlini) black hole, as it has to be.  We can also estimate the Kretschmann scalar for large distances ($r\rightarrow\infty$) and as a result we obtain:
\begin{equation}
R_{\mu\nu\rho\sigma}R^{\mu\nu\rho\sigma}\simeq\frac{2\L^2}{(n-1)^2(n-\al^2)^2}\left(2(1-\al^2)^2+n(n-1)+2(n-1)(1+\al^2)^2\right)b^{4\gamma}r^{-4\gamma}.
\end{equation}
The written above expression goes to zero when $r\rightarrow\infty$, but in the limit when $\al=0$ it takes finite value and it can be checked easily that the corresponding relation for AdS case is recovered.


\section{Thermodynamics of the black hole}
\begin{figure}
\centerline{\includegraphics[scale=0.3,clip]{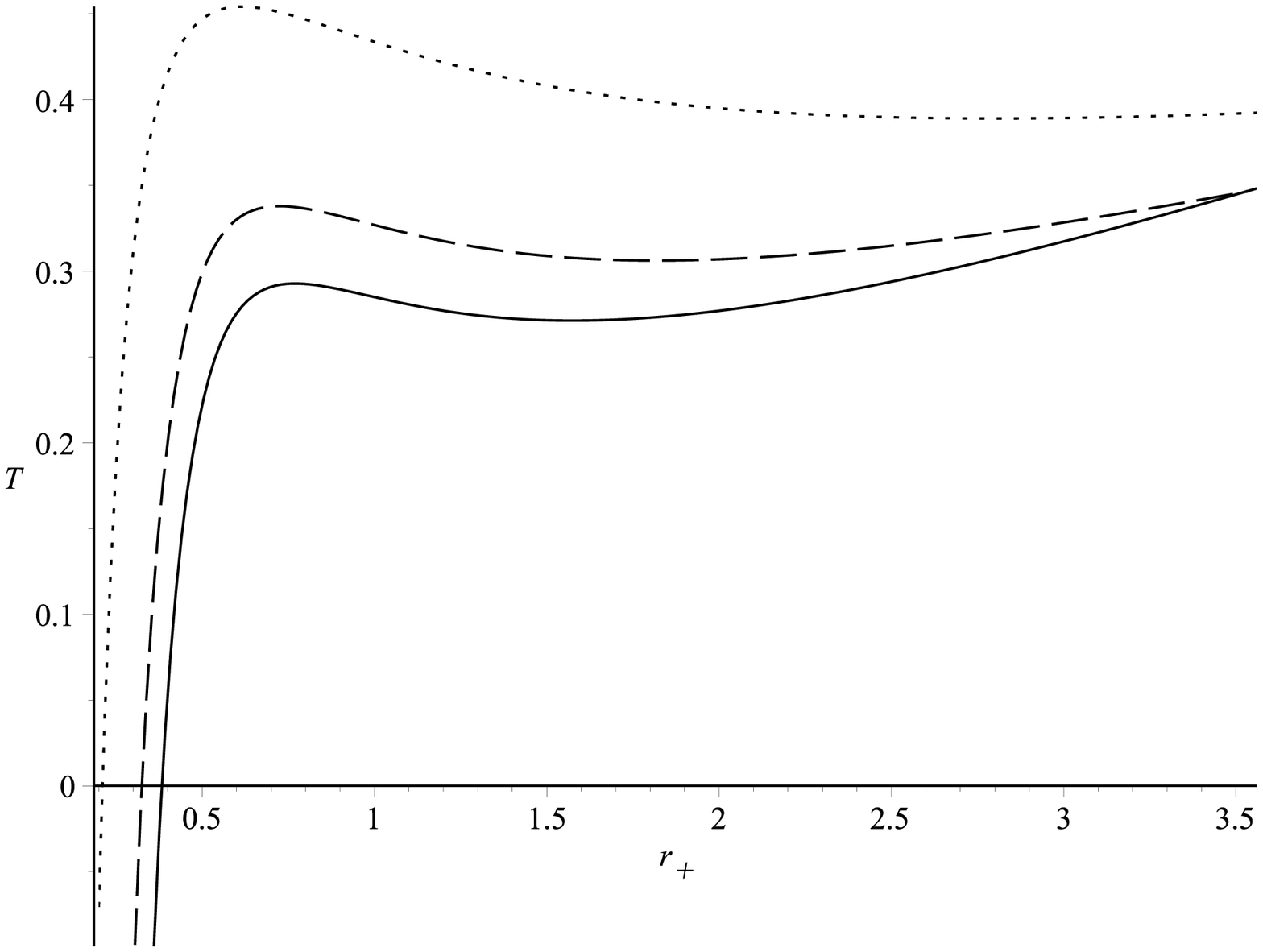}\includegraphics[scale=0.3,clip]{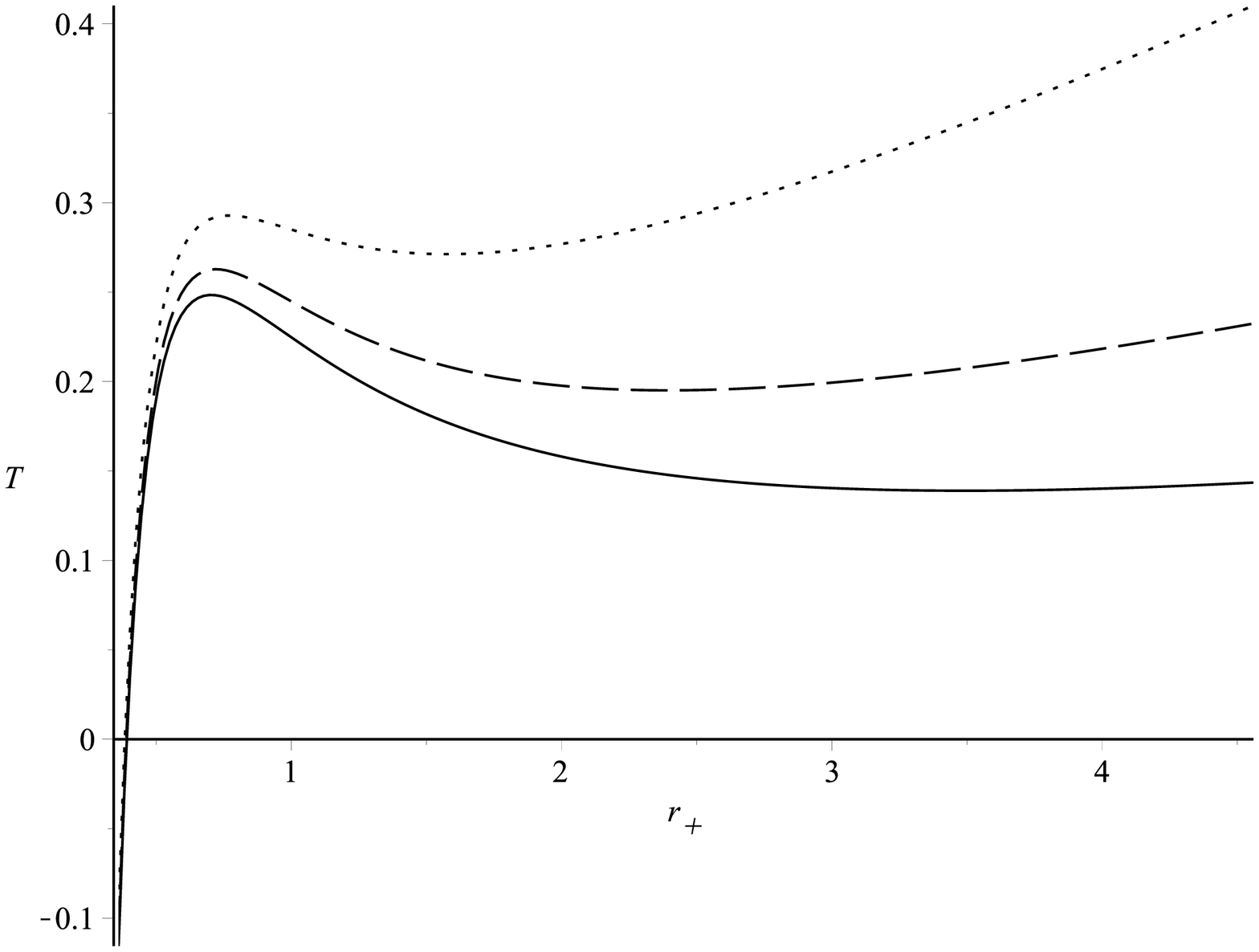}}
\caption{Black hole's temperature $T$ as a function of horizon radius  $r_+$ for several values of parameter $\al$ (the left graph) and cosmological constant $\Lambda$ (the right one). For both graphs we have taken: $n=5,b=1,q=0.4$. For the left plot the parameter $\al$ is chosen to be: $\al=0.1$, $\al=0.3$, $\al=0.5$ for solid, dashed and dotted curves respectively and $\L=-4$. For the right graph the varied parameter $\L$ is taken to be $\L=-1$, $\L=-2$ and $\L=-4$ for solid, dashed and dotted lines correspondingly, while $\al=0.1$.}\label{temp_graph}
\end{figure}

To investigate thermodynamics of the black we start from the derivation of the temperature, which is defined by virtue of surface gravity and we write it in the form: 
\begin{equation}
\kappa^2=-\frac{1}{2}\nabla_a\chi_b\nabla^a\chi^b, 
\end{equation}
where $\chi^{a}$ is a Killing vector, which is null on the horizon. Since we consider the static solution the Killing vector can be taken to be the time translation vector $\chi^{a}=\partial/\partial t$. Using the written above relation for the surface gravity we can write the relation for the temperature in the following form:
\begin{eqnarray}\label{temp}
T=\frac{\kappa}{2\pi}=\frac{(1+\al^2)}{4\pi}\left(\frac{n-2}{1-\al^2}b^{-2\gamma}r^{2\gamma-1}_{+}-\frac{\Lambda}{n-1}b^{2\gamma}r_+^{1-2\gamma}-\frac{n-2}{1-\al^2}q^2b^{-6\gamma}r_+^{6\gamma-3}\right), 
\end{eqnarray}
where $r_+$ denotes the radius of the event horizon of the black hole. The character of the obtained relation for temperature is rather complicated, but several quite general conclusions can be made from the formula (\ref{temp}). Firstly, we would like to describe the behaviour of the temperature for small as well as large values of the  horizon radii $r_+$,  obviously it is defined by the parameters given in the relation (\ref{temp}). Here and in the following we assume that the effective cosmological constant $\L$ is negative to avoid the existence of cosmological horizon and we also suppose that the coupling constant varies in the interval $0<\al<1$. Taking into account given above assumptions we can conclude that for large $r_+$ the dominant term in the relation (\ref{temp}) is the second one and in this case the temperature increases if the horizon radius $r_+$ goes up.  For small distances the character of dependence $T(r_+)$ is determined mainly by the interrelation between the first and the third terms. For better understanding of the function $T(r_+)$ we give graphical representation here, namely the Figure [\ref{temp_graph}]. Looking at these figures one can conclude that the variation of the coupling constant $\al$ affects considerably on the temperature for all values of $r_+$, whereas the variation of the cosmological constant $\L$ has substantial influence for large radii of the horizon $r_+$.  The above conclusions are rather expectable, because the parameter $\al$ is present in all the terms in the expression (\ref{temp}), whereas the cosmological constant $\L$ is present in the only term, which according to our assumption about the value of the parameter $\al$, determines the behaviour of the temperature for large horizon radius. We also point out that the temperature $T(r_+)$ might be non monotonous. This non monotonous behaviour gives us hint about some kind of critical behaviour and it will be studied below. 
 
Another important thermodynamic entity is the entropy which is defined in a standard way and equals to a quarter of the horizon's area, so in our case we write:
\begin{equation}\label{entropy}
S=\frac{\omega_{n-1}}{4}b^{\gamma(n-1)}r^{(n-1)(1-\gamma)}_+.
\end{equation}
Using the written relations for the entropy and temperature and taking into account the first law of black hole thermodynamics which should take the form:
\begin{equation}\label{first_law}
dM=TdS,
\end{equation}
where now:
\begin{equation}\label{temp_TD}
T=\left(\frac{\partial M}{\partial S}\right)_{\L,q},
\end{equation}
we can obtain the mass of the black hole, which can be written as follows:
\begin{equation}\label{mass_bh}
M=\frac{(n-1)b^{(n-1)\gamma}\omega_{n-1}}{16\pi(1+\al^2)}m,
\end{equation}
where $\omega_{n-1}$ is the surface area of a $n-1$--dimensional unit hypersphere. From the latter relation we see that the mass here is completely defined by the integration constant $m$  and coincides with corresponding relation obtained for Einstein-Maxwell-dilaton black holes \cite{Stetsko_EPJC19}. We also point out here that in case of Einstein-Maxwell-dilaton black holes the mass was derived by virtue of quasi-local approach \cite{Brown_PRD93}. Similar task can be performed here, but in contrast with the Einstein-Maxwell-dilaton case here we should take another reference background, namely in the reference metric it might be necessary to take into account the term corresponding to the Yang-Mills field, whereas in case of Einstein-Maxwell-dilaton theory the reference background was related only to the dilaton part.   

The relation (\ref{mass_bh}) can be rewritten as a completely thermodynamic relation, namely it can be represented as a function of the entropy: 
\begin{eqnarray}\label{mass_TD}
\nonumber M=\frac{(n-1)(1+\al^2)b^{(n-1)\gamma}\omega_{n-1}}{16\pi}\left(\frac{(n-2)b^{-2\gamma}}{(1-\al^2)(\al^2+n-2)}b^{-2\gamma}\left(\frac{4S}{\omega_{n-1}b^{(n-1)\gamma}}\right)^{\frac{n-2-(n-3)\gamma}{(n-1)(1-\gamma)}}-\right.\\\left.\frac{\Lambda}{(n-1)(n-\alpha^2)}b^{2\gamma}\left(\frac{4S}{\omega_{n-1}b^{(n-1)\gamma}}\right)^{\frac{n-(n+1)\gamma}{(n-1)(1-\gamma)}}+\frac{(n-2)q^2}{(\al^2-1)(n+3\alpha^2-4)}b^{-6\gamma}\left(\frac{4S}{\omega_{n-1}b^{(n-1)\gamma}}\right)^{\frac{(n-1)(1-\gamma)+6\gamma-3}{(n-1)(1-\gamma)}}\right).
\end{eqnarray} 

\begin{figure}
\centerline{\includegraphics[scale=0.33,clip]{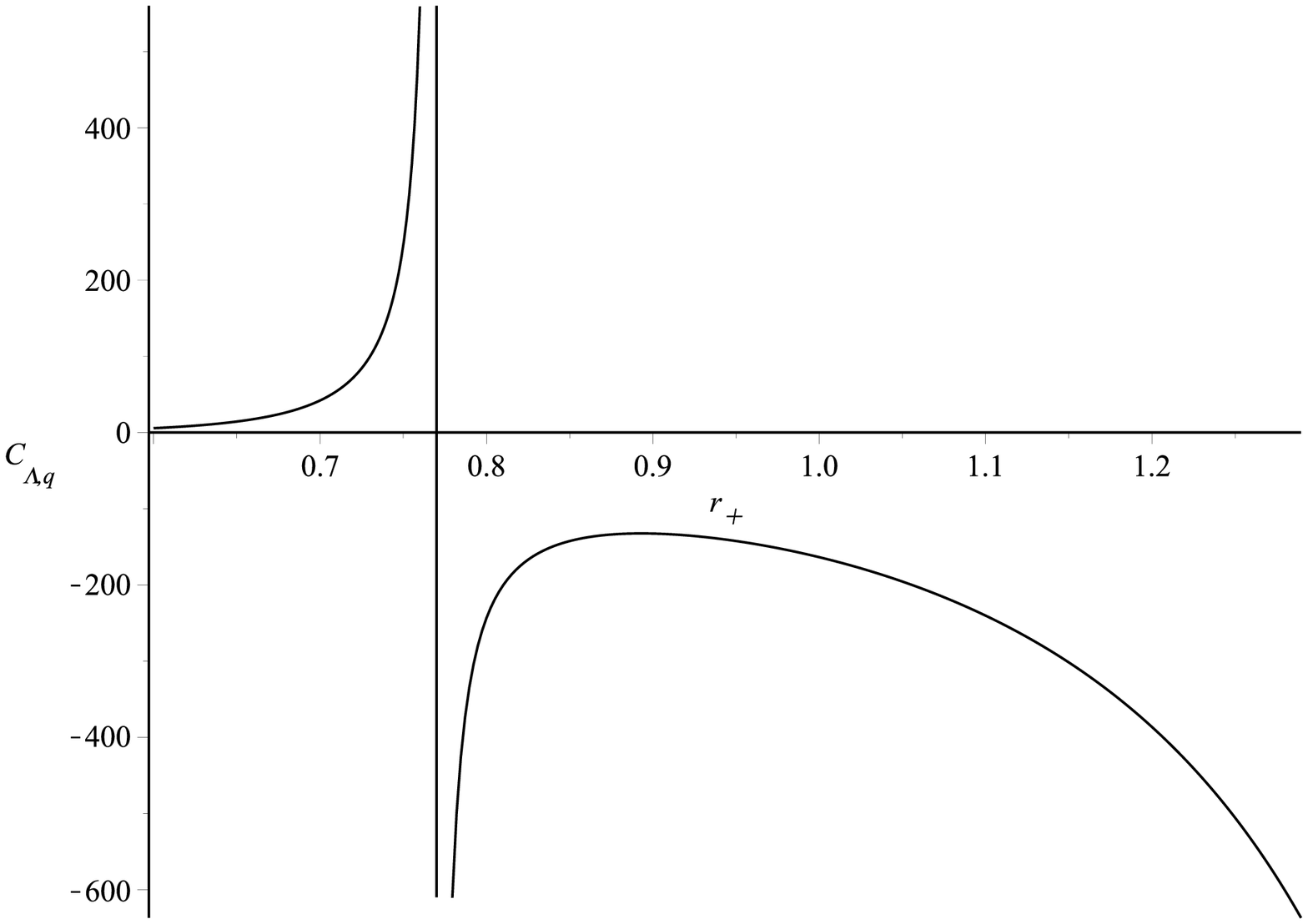}\includegraphics[scale=0.33,clip]{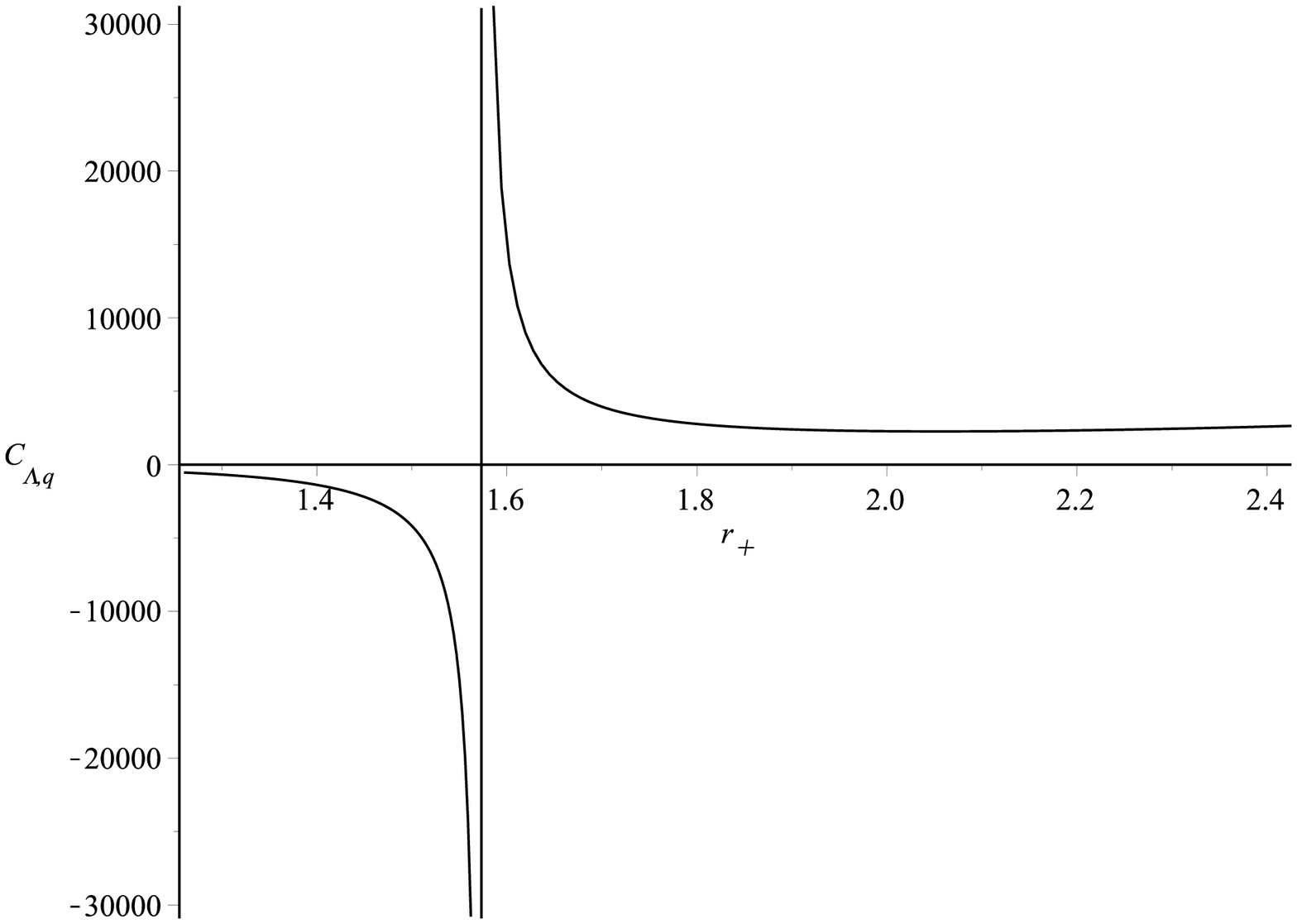}}
\caption{Heat capacity of the black hole $C_{\L,q}$ as a function of horizon radius  $r_+$  which demonstrates two discontinuity points. The fixed parameters are as follows: $n=5$, $\al=0.1$, $\L=-4$, $b=1$ and $q=0.4$.}\label{heatcap_gr}
\end{figure}

To study thermal stability of the black hole we calculate heat capacity, which is defined as follows:
\begin{equation}\label{heat_capacity}
C_{\L,q}=T\left(\frac{\partial S}{\partial r_+}\right)_{\L,q}\left(\frac{\partial T}{\partial r_+}\right)^{-1}_{\L,q}
\end{equation}
Having made some elementary calculations we arrive at the expression:
\begin{eqnarray}
\nonumber C_{\L,q}=\frac{(n-1)\omega_{n-1}}{4}b^{\gamma(n-1)}r^{(n-1)(1-\gamma)}_+\left(\frac{n-2}{1-\al^2}b^{-2\gamma}r^{2\gamma-1}_+-\frac{\Lambda}{n-1}b^{2\gamma}r_+^{1-2\gamma}-\frac{n-2}{1-\al^2}q^2b^{-6\gamma}r_+^{3(2\gamma-1)}\right)\\\times\left(-(n-2)b^{-2\gamma}r^{2\gamma-1}_{+}-\frac{\Lambda(1-\alpha^2)}{n-1}b^{2\gamma}r_+^{1-2\gamma}-{3(n-2)}q^2b^{-6\gamma}r_+^{2(3\gamma-2)}\right)^{-1}.\label{heat_capac}
\end{eqnarray} 
Since the temperature (\ref{temp}) is not a monotonous function and as it was demonstrated graphically (see Fig.[\ref{temp_graph}]) it might have two extrema points what gives rise to the conclusion that the heat capacity $C_{\L,q}$ might be discontinuous. The points of discontinuity separate thermodynamically stable and unstable regions. We can also conclude that with increasing of the cosmological constant $\L$ in absolute value the nonmonotonicity of the temperature becomes weaker and with the following increase it might disappear, and it means that in this case the heat capacity $C_{\L,q}$ turns to be a continuous function. The behavior of the heat capacity $C_{\L,q}$ as a function of the horizon radius $r_+$ is demonstrated on the Figure [\ref{heatcap_gr}]. The graphs depicted on the Figure [\ref{heatcap_gr}] corroborate the conclusions made above, in fact there are two points of discontinuity, namely for large horizon radius $r_+$ the black hole is stable, since the heat capacity is positive $C_{\L,q}$, then there a discontinuity point which corresponds to an extreme point of the temperature which is followed by unstable domain for intermediate values of $r_+$ and finally we have the second discontinuity point and the stable domain for small horizon radius. It should be pointed out that for very small radius of horizon the heat capacity again becomes negative but here we have continuous transition from positive to negative values, this fact can be easily explained by virtue of the relation (\ref{heat_capacity}), namely both of the derivatives in this relation are positive for these values of $r_+$ whereas the temperature turns to be negative, thus the heat capacity also takes negative values.

\begin{figure}
\centerline{\includegraphics[scale=0.3,clip]{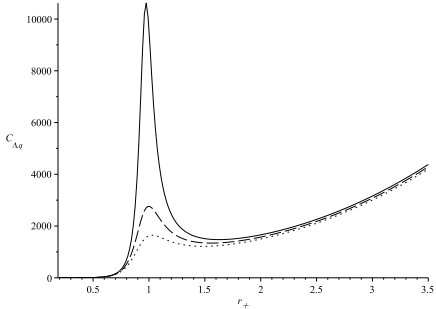}}
\caption{Heat capacity $C_{\L,q}$ as a function of horizon radius $r_+$ after disappearance of discontinuity. The correspondence of lines is as follows: the solid curve corresponds to $\L=-8.4$, the dashed curve corresponds to $\Lambda=-8.6$ and the dotted line corresponds to $\Lambda=-8.8$. The other fixed parameters are as follows: $n=5$, $\al=0.5$, $b=1$ and $q=0.4$.}\label{heat_cap_cont}
\end{figure}

If the cosmological constant $\L$ increases in absolute value the discontinuity points are getting closer and finally they merge into one point with the following transformation of the discontinuity into a peak of finite height. With further increase of the module of $\L$ the height of the peak decreases and then it disappears. This behaviour is demonstrated on the Figure [\ref{heat_cap_cont}]. Finally, if the module of the cosmological constant $\L$ is large enough the heat capacity might transform into a monotonous function which is positive for all the values of the horizon radius larger than the mentioned above   point when the temperature becomes negative, and as a consequence the system becomes thermodynamically stable for all these values of the horizon radius. We note that similar behaviour of heat capacity takes place for Einstein-Maxwell-dilaton black hole \cite{Stetsko_EPJC19}.  

\section{Thermodynamics in extended phase space}
In contrast to the previous section where the cosmological constant was assumed to be fixed here we suppose that it might be varied \cite{Kastor_CQG09, Cvetic_PRD11,Dolan_CQG11}. This assumption has far reaching  consequences since it allows to develop very close ties with many aspects of thermodynamics of condensed matter systems that were beyond the scope before. Namely, it gives opportunity to apply well-established techniques related to critical phenomena to characterize thermodynamic behaviour of the black holes \cite{Kubiznak_JHEP12,Kubiznak_CQG17}. If the cosmological constant is supposed to be a thermodynamic variable it is identified with thermodynamic pressure \cite{Kastor_CQG09}. Since the new thermodynamic value is introduced it means that the black hole's mass should not be considered as the internal energy as it was in the previous section, but now it is identified with the enthalpy function. The extension of the thermodynamic phase space allows to derive an equation of state which is an analog of the well-known Van der Waals equation of state. In addition, in the extended space one can establish Smarr relation what is not always possible in the standard framework.

In our case we introduce thermodynamic pressure similarly as it was done for other dilaton black holes \cite{Stetsko_EPJC19}:
\begin{equation}\label{press}
P=-\frac{\Lambda}{16\pi}\left(\frac{b}{r_+}\right)^{2\gamma}.
\end{equation}
We have mentioned above that now the black hole's mass should be identified with the enthalpy $M=H$, taking this fact into account we derive thermodynamic volume by virtue of well-known thermodynamic relation:
\begin{equation}\label{TD_volume}
V=\left(\frac{\partial H}{\partial P}\right)_S=\left(\frac{\partial M}{\partial P}\right)_S.
\end{equation} 
Having used of the latter relation we obtain the evident form for the thermodynamic volume:
\begin{equation}\label{td_vol}
V=\frac{\omega_{n-1}(1+\alpha^2)}{n-\alpha^2}b^{(n-1)\gamma}r_+^{(n-1)(1-\gamma)+1}.
\end{equation} 
It can be checked easily that in the limit $\al=0$ the latter relation leads  to the volume of a ball with the radius equal to the horizon radius $r_+$.

It is known that for Einstein-Maxwell-dilaton black holes the total charge of the black holes can be treated as a thermodynamic value. In our case of Einstein-Yang-Mills-dilaton theory there are some peculiarities since the charge related to the Yang-Mills field is defined in a different way in comparison with the Maxwell field case. We introduce Yang-Mills charge in the following way:
\begin{equation}
Q=\frac{1}{4\pi\sqrt{(n-1)(n-2)}}\int_{\Sigma}d^{n-1}\chi J(\Omega)\sqrt{Tr(F^{(a)}_{\mu\nu}F^{(a)}_{\mu\nu})}=\frac{\omega_{n-1}}{4\pi}q.
\end{equation} 
This integral should be taken over angular variables of a sphere that encloses the black hole and in the written above integral $J(\Omega)$ denotes corresponding Jacobian for the spherical variables.
Similarly to the given above thermodynamic volume $V$ as a conjugate to the pressure $P$ we introduce Yang-Mills potential as a conjugate to the charge $Q$, namely we can write:
\begin{equation}
U=\left(\frac{\partial M}{\partial Q}\right)_{S,P}.
\end{equation}
Having used the defined above thermodynamic values, namely the pressure $P$, the Yang-Mills charge $Q$ and their conjugates we can write the extended first law for the black hole:
\begin{equation}\label{ext_first_law}
dM=TdS+VdP+UdQ.
\end{equation}
It has been pointed out above that within the extended thermodynamics one can derive the Smarr relation which shows the relation between all the thermodynamic values. In our case the Smarr relation takes the form:
\begin{equation}\label{smarr_gen}
(n+\al^2-2)M=(n-1)TS+2(\al^2-1)VP+(1-\al^2)UQ.
\end{equation}
In the limit when $\al=0$, the latter equation can be rewritten in the form:
\begin{equation}\label{smarr_simpl}
(n-2)M=(n-1)TS-2VP+UQ.
\end{equation}
It is easy to see that the factors in the left hand side of the written equation as well as near the $TS$ and $VP$ terms are the same as for other types of the black holes in the framework of standard General Relativity \cite{Kubiznak_CQG17}.

\begin{figure}
\centerline{\includegraphics[scale=0.33,clip]{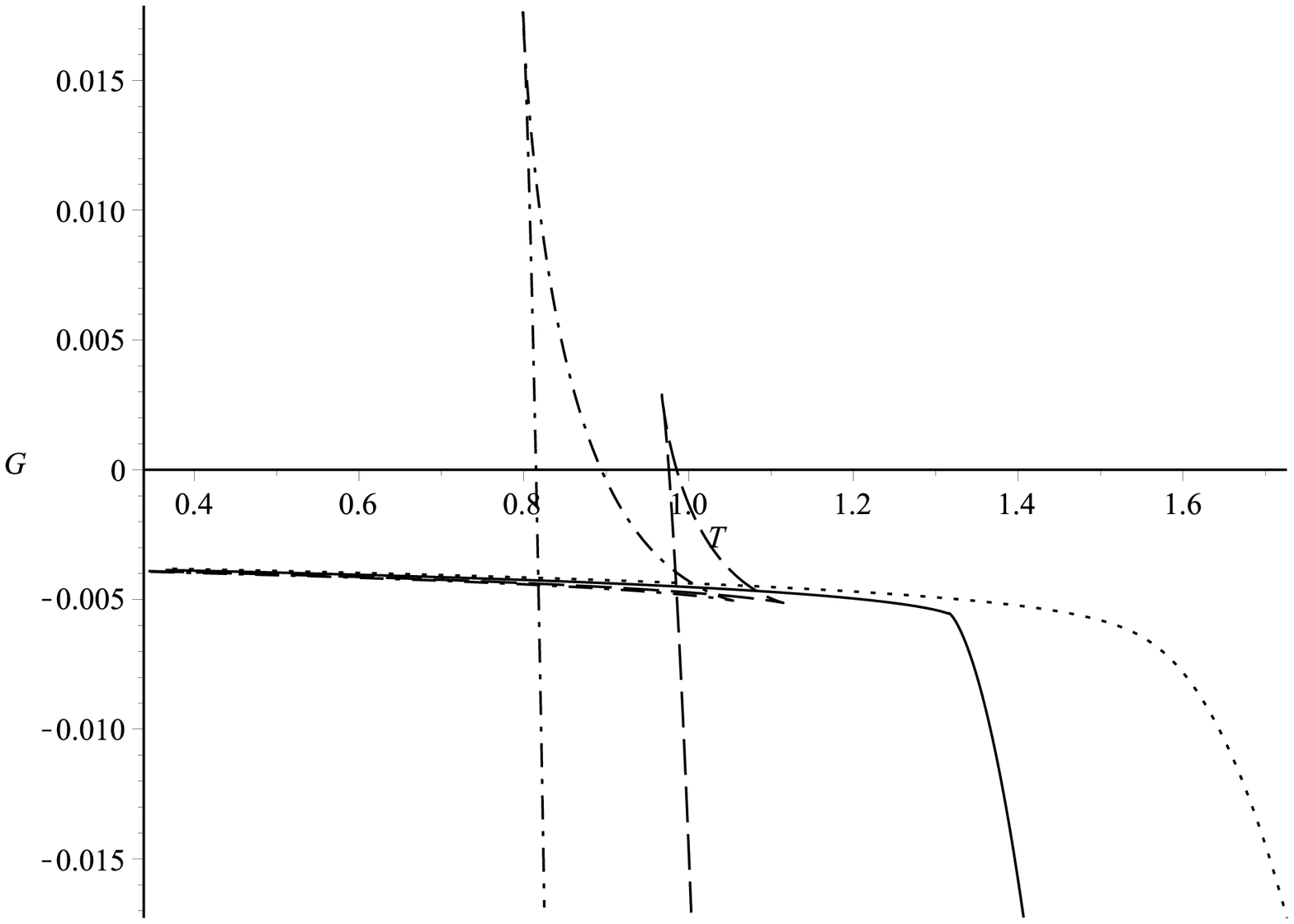}\includegraphics[scale=0.33,clip]{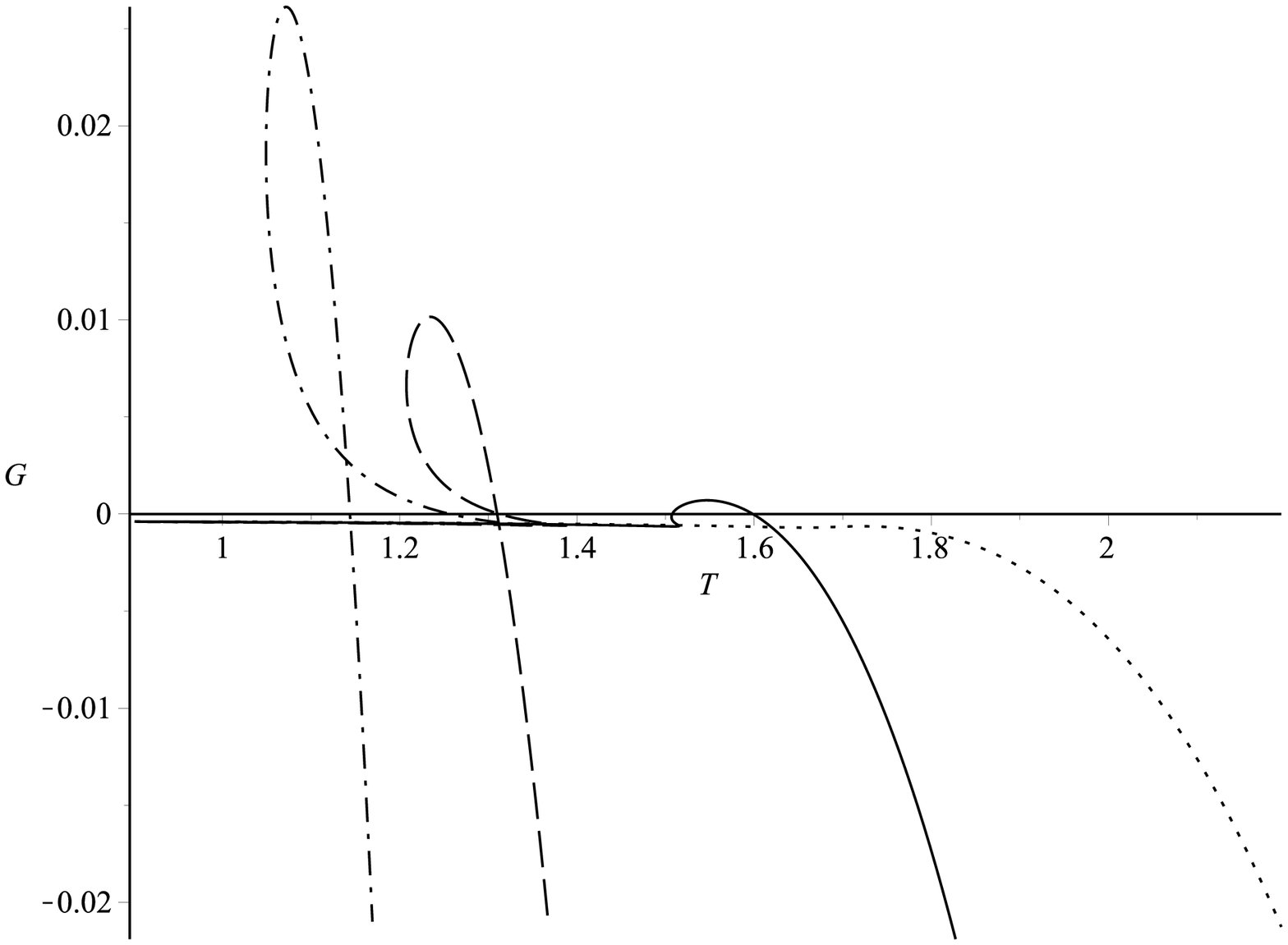}}
\caption{Gibbs free energy $G$ as a function of temperature for fixed values of pressure. For both graphs we have $n=5$, $b=1$ and correspondence of the lines is the following the dotted, solid, dashed and dash-dotted lines correspond to the pressures $P=3/2P_c$, $P=P_c$, $P=P_c/2$ and $P=P_c/3$ respectively and here $P_c$ denotes the critical pressure, which is defined for given values of $n$, $\al$, $b$ and $q$. We also note that for the left graph we have $\al=0.1$ and $q=0.1$, whereas for the right graph $\al=0.5$ and $q=0.12$.}\label{gibbs_gr}
\end{figure}

\begin{figure}[!]
\centerline{\includegraphics[scale=0.33,clip]{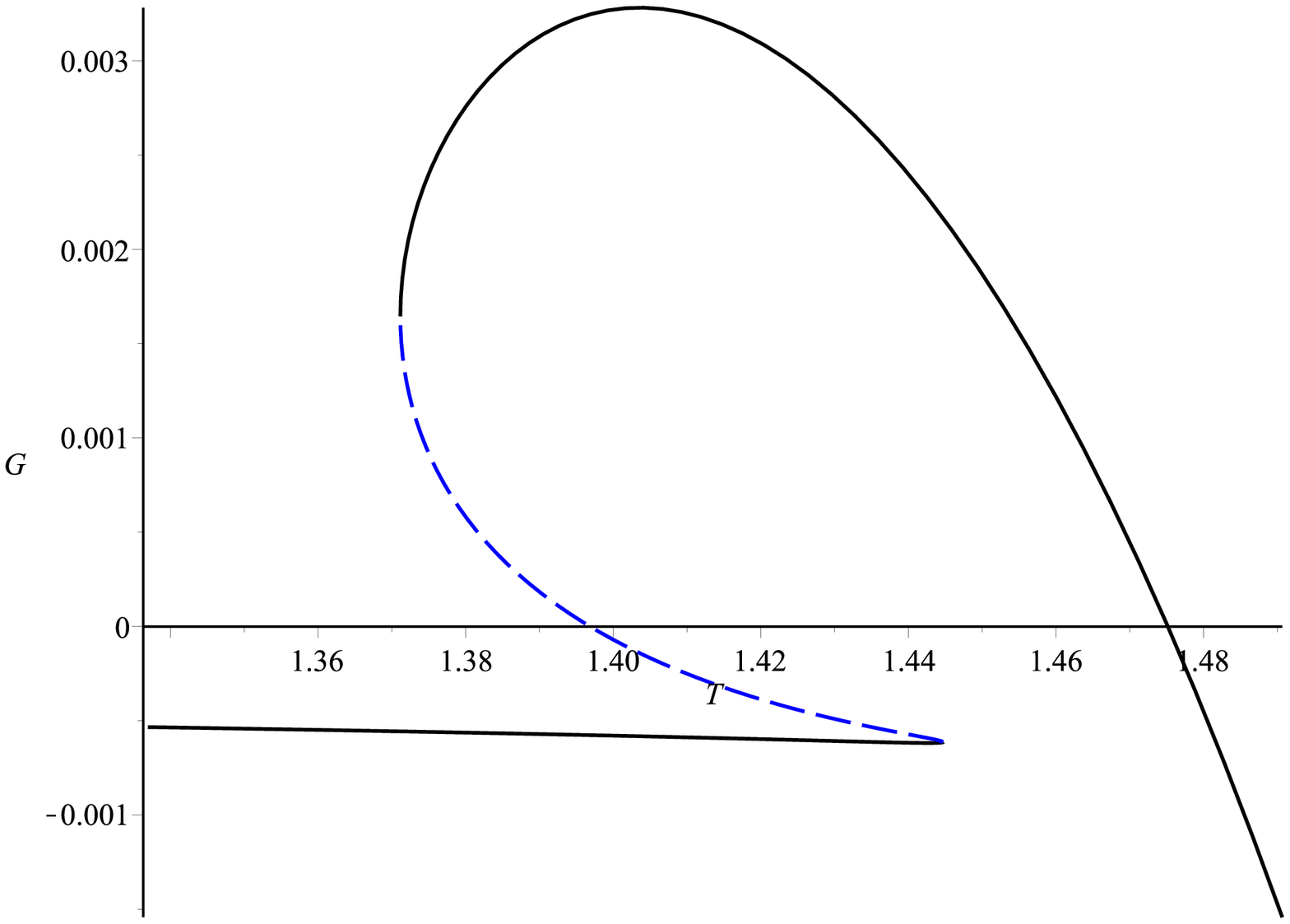}\includegraphics[scale=0.33,clip]{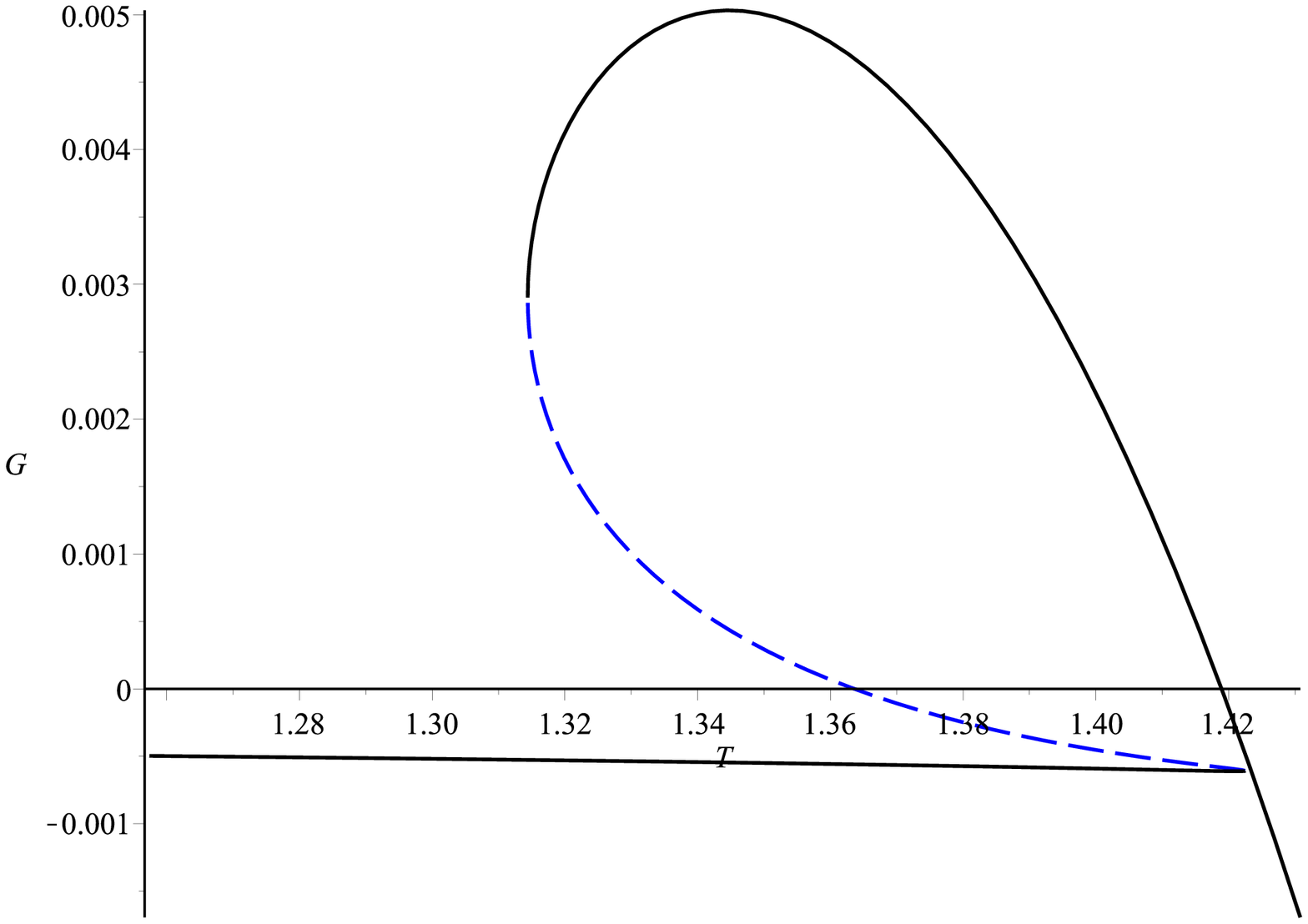}}
\caption{Gibbs free energy $G$ as a function of temperature $T$ which shows instability region given by dashed parts of the curves.}\label{gibbs_inst}
\end{figure}

The relations (\ref{temp}), (\ref{press}) and (\ref{TD_volume}) allows one to obtain the equation of state which can be considered as an analog of the Van der Waals equation \cite{Kubiznak_JHEP12,Kubiznak_CQG17}. For convenience we do not express the pressure as an evident function of the thermodynamic volume $V$ but keep the dependence of the horizon radius $r_+$. Thus, we write the equation of state in the following form:
\begin{eqnarray}\label{eq_of_state}
P=\frac{(n-1)}{4(1+\alpha^2)}\frac{T}{r_+}-\frac{(n-1)(n-2)}{16\pi(1-\alpha^2)}b^{-2\gamma}r_+^{2(\gamma-1)}\left(1-q^2b^{-4\gamma}r^{2(2\gamma-1)}_+\right).
\end{eqnarray}
To establish a closer relation between the written above equation of state (\ref{eq_of_state}) and the Van der Waals equation of state instead of our ``geometrical'' pressure $P$ and the temperature $T$ we can introduce  the ``physical'' pressure and temperature as follows \cite{Kubiznak_JHEP12}:
\begin{equation}
[P]=\frac{\hbar c}{l^{n-1}_{Pl}}P, \quad [T]=\frac{\hbar c}{k}T,
\end{equation}
where $l_{Pl}$ is the Planck length in $n$--dimensional space and $k$ is the Boltzmann constant. It should be pointed out that after redefinition of the thermodynamic quantities a new specific volume appears in the equation of state (\ref{eq_of_state}), this volume is proportional to the product of the horizon radius $r_+$ over $l^{n-1}_{Pl}$. Keeping the geometrical units we can rewrite the  equation of state (\ref{eq_of_state}) in the form:
\begin{eqnarray}\label{eq_of_st_2}
P=\frac{T}{v}-\frac{(n-2)(1+\alpha^2)}{4\pi(1-\alpha^2)}\kappa^{2\gamma-1}b^{-2\gamma}v^{2(\gamma-1)}\left(1-q^2b^{-4\gamma}\kappa^{2(2\gamma-1)}v^{2(2\gamma-1)}\right),
\end{eqnarray}
where we have introduced new specific ``volume'' $v$ of the form:
\begin{equation}\label{sp_vol}
v=\frac{4(1+\alpha^2)}{n-1}r_+
\end{equation}
and in the upper equation of state $\kappa=(n-1)/(4(1+\alpha^2))$. The rewritten equation of state (\ref{eq_of_st_2}) is now treated as the analog of the Van der Waals equation and it can be examined in a similar way. Namely, what is interesting for us is the critical behaviour of the black hole and appearance of a phase transition which takes place between the so called small and large black holes as it is for ordinary charged or Einstein-Maxwell-dilaton  balck holes \cite{Kubiznak_JHEP12,Kubiznak_CQG17,Stetsko_EPJC19}. To investigate this issue one should find an inflection point which is defined as follows:
\begin{equation}
\left(\frac{\partial P}{\partial v}\right)_T=0, \quad \left(\frac{\partial^2 P}{\partial v^2}\right)_T=0.
\end{equation}  
Using the latter relations we obtain critical specific volume $v_c$ and critical temperature $T_c$: 
\begin{equation}\label{v_c}
v_c=\frac{4(1+\al^2)}{n-1}\left(3(2-\al^2)q^2b^{-4\gamma}\right)^{\frac{1}{2(1-2\gamma)}},
\end{equation}
\begin{equation}\label{t_c}
T_c=\frac{(n-1)(n-2)}{12\pi(1-\alpha^2)(1+\alpha^2)}\kappa^{2(\gamma-1)}b^{-2\gamma}v_c^{2\gamma-1}.
\end{equation}
As a consequence, for the critical pressure we arrive at:
\begin{equation}\label{p_c}
P_c=\frac{(n-1)(n-2)(1-\al^2)}{16\pi(1+\alpha^2)(2-\al^2)}\kappa^{2(\gamma-1)}b^{-2\gamma}v_c^{2(\gamma-1)}.
\end{equation}
From the written above relations we obtain critical ratio in the form:
\begin{equation}\label{cr_rat}
\rho_c=\frac{P_cv_c}{T_c}=\frac{3(1-\alpha^2)^2}{4(2-\alpha^2)}.
\end{equation}
It should be pointed out here that the critical ratio does not depend on the dimension $n$ and the limit $\alpha=0$ the well-known critical ratio for Van der Waals gas in $3$-dimensional space is recovered (for arbitrary dimension of black hole's space-time):
\begin{equation}
\left.\frac{P_cv_c}{T_c}\right|_{\alpha=0}=\frac{3}{8}.
\end{equation}
Having performed a  Legendre transformation for the enthalpy we obtain the Gibbs free energy:
\begin{eqnarray}\label{Gibbs_pot}
\nonumber G(T,P)=\frac{\omega_{n-1}(1+\alpha^2)b^{(n-1)\gamma}}{16\pi}r_+^{(n-1)(1-\gamma)}\left(\frac{n-2}{\alpha^2+n-2}b^{-2\gamma}r_+^{2\gamma-1}-\right.\\\left.\frac{16\pi(1-\alpha^2)P}{(n-1)(n-\alpha^2)}r_{+}-\frac{3(n-2)}{(n+3\alpha^2-4)}q^2b^{-6\gamma}r_{+}^{3(2\gamma-1)}\right).
\end{eqnarray} 
The dependence $G=G(T)$ for some fixed values of the pressure $P$ is shown graphically on the Figure [\ref{gibbs_gr}]. The left graph on the Figure [\ref{gibbs_gr}] shows the specific swallow-tail behaviour of the Gibbs free energy for the pressures below the critical one, when the coupling constant is relatively small ($\al=0.1$), and when the pressure equals to the critical value $P_c$ the Gibbs free energy is piecewise smooth function. The swallow-tail behaviour of the Gibbs free energy is known to take place for Van der Waals systems (for instance a liquid-gas system) in condensed matter systems, this kind of behaviour also appears in the framework of extended thermodynamics for RN-AdS black hole \cite{Kubiznak_JHEP12,Kubiznak_CQG17} or for Einstein-Maxwell-dilaton black holes \cite{Stetsko_EPJC19}. If the parameter $\al$ increases ($\al=0.5$) additional feature appears for the pressures below and equal to the critical one, namely for the critical pressure $P_c$ we have specific maximum at the critical point ($T=T_c$) and for the pressures below the critical one there is specific loop forming domain, when the swallow-tail is not yet formed, and the following decrease of the pressure gives rise to the appearance of the swallow-tail, but with additional closed loop (see Figure [\ref{gibbs_inst}]). Similar behaviour of the Gibbs free energy for various values of the parameter $\al$ takes place for the Einstein-Maxwell-dilaton black hole \cite{Dehyadegari_PRD17,Stetsko_EPJC19}. It was shown that in addition to the first order phase transition which happens for the Van der Waals systems the dilaton field gives rise to the appearance of the zeroth order phase transition \cite{Dehyadegari_PRD17,Stetsko_EPJC19} when the Gibbs free energy becomes discontinuous.

To investigate the domain of instability which is related to the swallow-tail behaviour we introduce isothermal compressibility which is defined as follows:
\begin{equation}\label{iso_comp}
\kappa_T=-\frac{1}{V}\left(\frac{\partial V}{\partial P}\right)_T.
\end{equation}
Since our system fulfills the equation of state (\ref{eq_of_st_2}) we obtain the evident form for the isothermal compressibility:
\begin{equation}\label{isot_compr}
\kappa_T=\frac{n+\alpha^2}{1+\alpha^2}\left(P-\frac{n-2}{4\pi}\kappa^{2\gamma-1}b^{-2\gamma}v^{2(\gamma-1)}\left(1-3q^2b^{-4\gamma}\kappa^{2(2\gamma-1)}v^{2(2\gamma-1)}\right)\right)^{-1}.
\end{equation}
We note here that the isothermal compressibility is considered as a function of the pressure $P$ and the temperature $T$. It can verified easily that for the isobars above the critical one ($P>P_c$) the isothermal compressibility $\kappa_T$ is positive for all temperatures and it gives  rise to the conclusion about the stability of the system in this range of pressures. Similar conclusion can be made if one examines the isotherms for the temperatures above the critical one $T>T_c$. At the critical point ($P=P_c$, $T=T_c$) there is a phase transition of the second order which takes place for the Van der Waals systems.

For the isobars below the critical one we have similarity with Einstein-Maxwell-dilaton black holes, namely when the closed loop is not formed yet there is a domain where the Gibbs free energy is discontinuous and there is the phase transition of the zeroth order in this domain \cite{Dehyadegari_PRD17,Stetsko_EPJC19}. The Figure [\ref{gibbs_inst}] demonstrates this discontinuity, namely on the left graph where the loop is not closed yet we have discontinuity of the Gibbs free energy, whereas on the right graph when the loop becomes closed the discontinuity of the Gibbs free energy disappears.

\begin{figure}[!]
\centerline{\includegraphics[scale=0.33,clip]{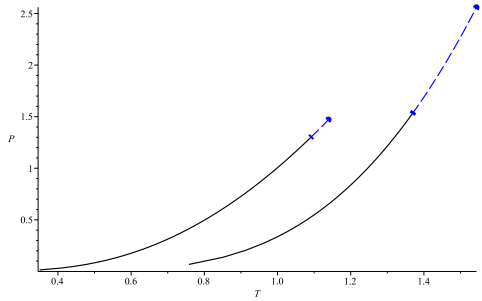}}
\caption{$P-T$ diagram (coexistence diagram) for two phases. Here the solid parts of the curves represent the domain where the first order phase transition takes places, whereas the dashed parts (which is just schematic to reflect the domain where the Gibbs free energy is discontinuous). For the both curves $n=5$, $b=1$, $q=0.12$, for the left curve $\al=0.1$, for the right one $\al=0.5$.}\label{coex_curves}
\end{figure}

After the closing of the loop when swallow-tail appears we have the phase transition of the first order and the Gibbs free energy is constant during the phase transition. This fact helps us to derive the coexistence curve for two phases (the so called small and large black holes), namely we make use of the Maxwell's equal areas law, which can be written in the form:
\begin{equation}\label{max_law}
P_0(v_2-v_1)=\int^{v_2}_{v_1}P{\rm d}v,
\end{equation}
where $v_1$ and $v_2$ denotes some volumes corresponding to the first (small) and the second (large) phases which are at the same pressure $P_0$. Using the equation (\ref{eq_of_st_2}) and after integration we obtain:
\begin{equation}\label{max_law_int}
P_0(v_2-v_1)=T\ln{\left(\frac{v_2}{v_1}\right)}+\frac{1+\alpha^2}{1-\alpha^2}A\left(v^{2\gamma-1}_2-v^{2\gamma-1}_1\right)-\frac{1+\alpha^2}{3(1-\alpha^2)}B\left(v^{3(2\gamma-1)}_2-v^{3(2\gamma-1)}_1\right),
\end{equation}
where we use the following notations:
\begin{equation}
A=\frac{(n-2)(1+\alpha^2)}{4\pi(1-\alpha^2)}\kappa^{2\gamma-1}b^{-2\gamma}, \quad B=\frac{(n-2)(1+\alpha^2)}{4\pi(1-\alpha^2)}\kappa^{3(2\gamma-1)}b^{-6\gamma}q^2.
\end{equation}
Using the equation of state (\ref{eq_of_st_2}) we can write:
\begin{equation}\label{temp_rel}
T\left(\frac{1}{v_2}-\frac{1}{v_1}\right)-A\left(v^{2(\gamma-1)}_2-v^{2(\gamma-1)}_1\right)+B\left(v^{2(3\gamma-2)}_2-v^{2(3\gamma-2)}_1\right)=0.
\end{equation}
Since we are going to find a coexistence curve we introduce a new parameter as a ratio of two volumes, namely $x=v_1/v_2$ and it is clear that $0<x<1$. Now we rewrite the latter relation in the following form:
\begin{equation}\label{temp_x}
T=\frac{x}{x-1}\left(Av^{2\gamma-1}_2\left(1-x^{2(\gamma-1)}\right)-Bv^{3(2\gamma-1)}_2\left(1-x^{2(3\gamma-2)}\right)\right).
\end{equation}
The volume of a phase (in our case the volume $v_2$) can be expressed as a function of the parameter $x$:
\begin{equation}\label{v_2x}
v^{2(2\gamma-1)}_2=\frac{A}{B}\frac{\left(\frac{2}{1-\al^2}\left(1-x^{2\gamma-1}\right)-\frac{x}{x-1}\ln{x}\left(1-x^{2(\gamma-1)}\right)\right)}{\left(\frac{2(2-\al^2)}{3(1-\al^2)}\left(1-x^{3(2\gamma-1)}\right)-\frac{x}{x-1}\ln{x}\left(1-x^{2(3\gamma-2)}\right)\right)}.
\end{equation}
We also use the relation (\ref{temp_x}) to rewrite the equation of state (\ref{eq_of_st_2}) in the following form:
\begin{equation}\label{press_x}
P=\frac{1}{x-1}\left(Av^{2(\gamma-1)}_2\left(1-x^{2\gamma-1}\right)-Bv^{2(3\gamma-2)}_2\left(1-x^{3(2\gamma-1)}\right)\right).
\end{equation}
It should be noted that when $x=1$ the relations (\ref{press_x}) and (\ref{temp_x}) give rise to the critical values for the pressure $P_c$ and the temperature previously obtained $T_c$.  Using the written above relations (\ref{temp_x}) and (\ref{press_x}) we can obtain coexistence relation ($P-T$--relation) for two phases where the first order phase transition takes place (see Figure [\ref{coex_curves}]). In general the character of $P-T$--dependence is similar to the corresponding dependence for the Einstein-Maxwell-dilaton black hole \cite{Stetsko_EPJC19}. Similarly to the Einstein-Maxwell-dilaton case when the parameter $\al$ increases the domain where the discontinuity  of the Gibbs free energy takes place widens. If the dilaton parameter $\al\rightarrow 0$ the domain with the discontinuity of the Gibbs free energy disappears.
The slope of the coexistence curve can be obtained by virtue of the Clapeyron equations which takes the form:
\begin{equation}\label{Clapeyron}
\frac{dP}{dT}=\frac{L}{T(v_2-v_1)},
\end{equation}
and here $L=T(s_2-s_1)$ is the latent heat of the phase change and and $s_i$ are the entropies of two phases which in our case represent the small and large black holes, and the latent heat shows gain or loss of the mass. The latter relation can be also used to find this latent heat, namely we rewrite the latter relation in the form:
\begin{equation}\label{latent_heat}
L=v_2(1-x)T(x)\frac{dP}{dx}\frac{dx}{dT}.
\end{equation}
\begin{figure}[!]
\centerline{\includegraphics[scale=0.33,clip]{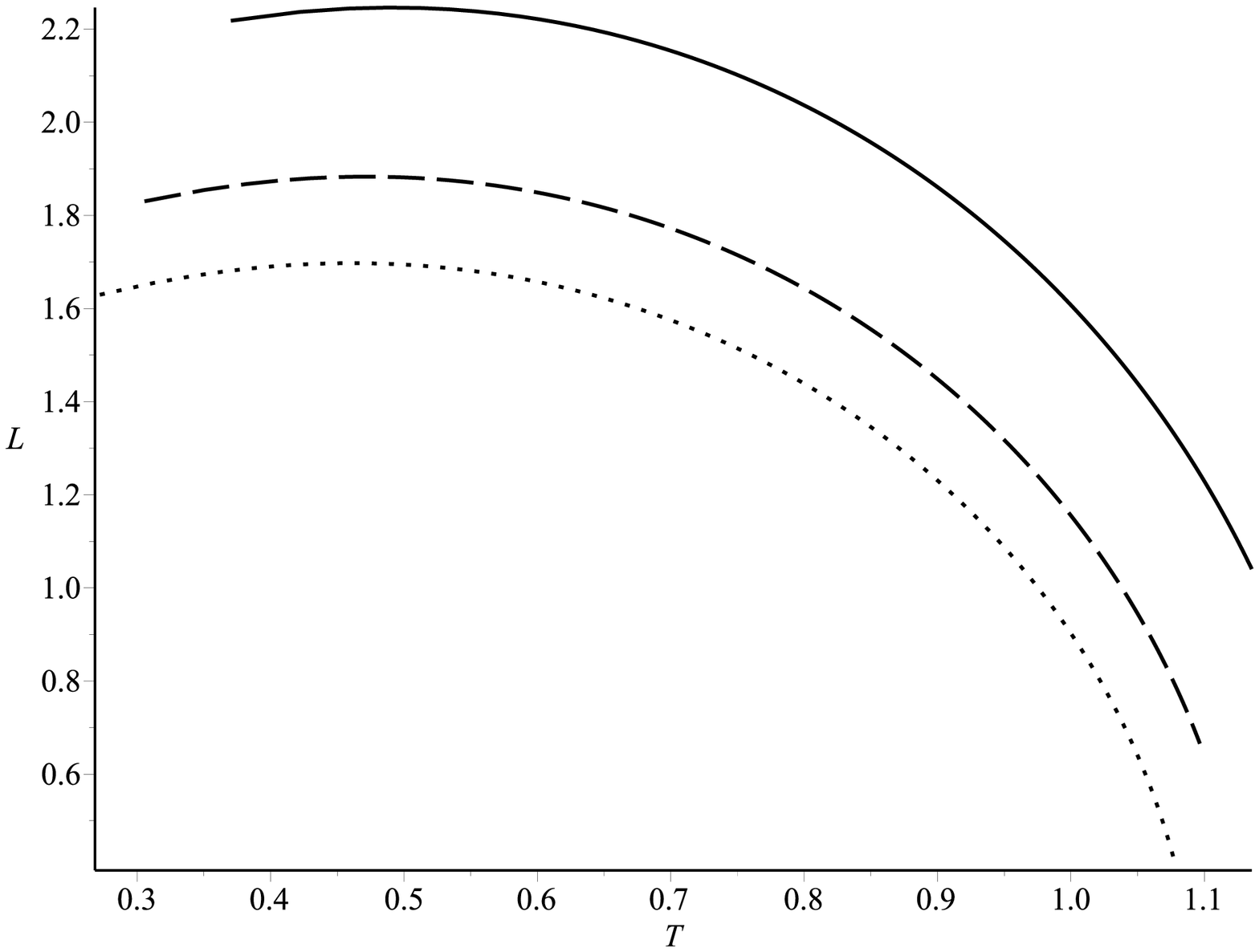}\includegraphics[scale=0.33,clip]{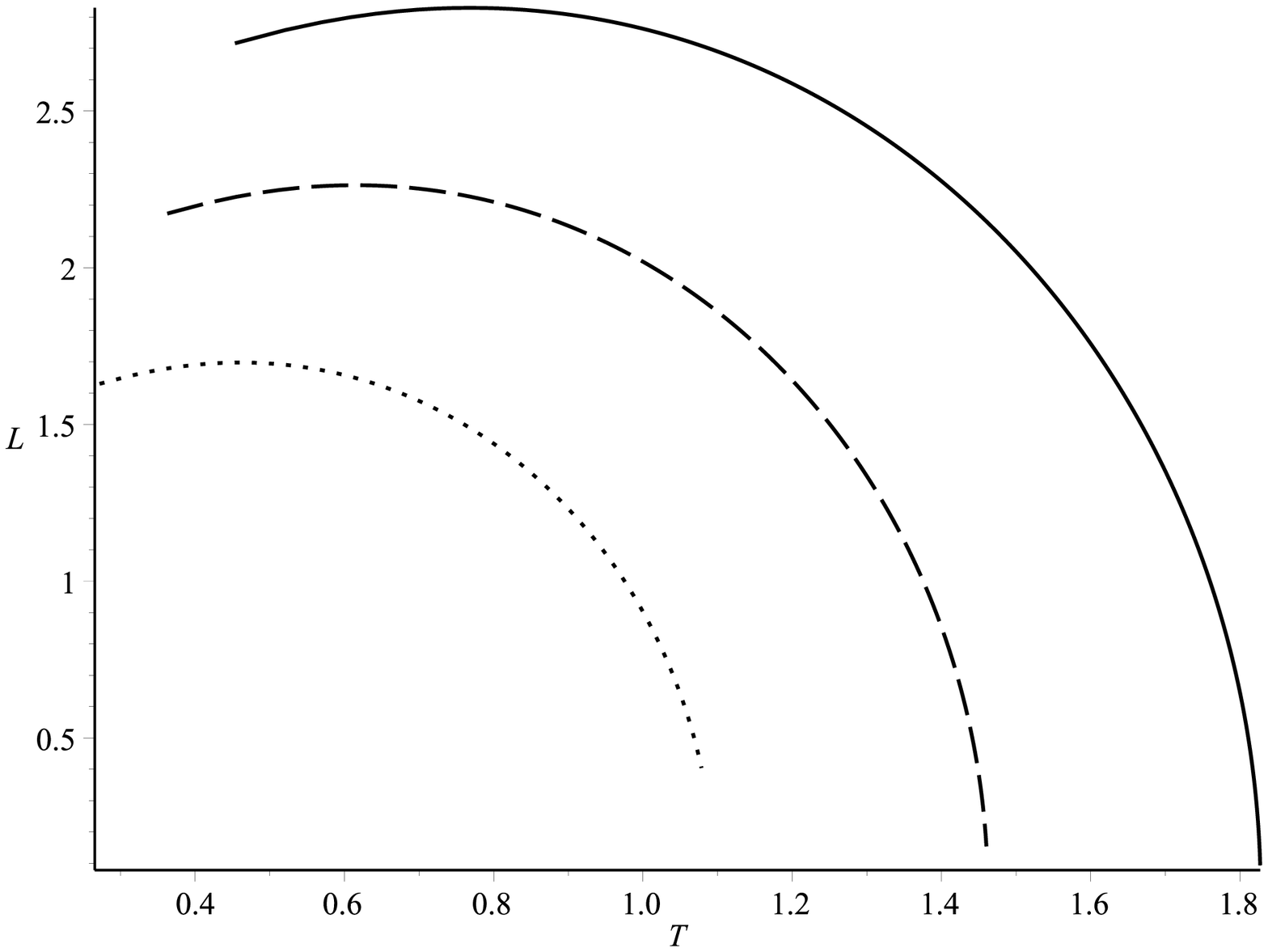}}
\caption{Latent heat $L$ as a function of temperature $T$, for several values of the parameter $\al$ (the left graph) and the dimension $n$ (the right graph). For the both graphs we have $b=1$ and $q=0.12$. For the left graphs the parameter $\al$ takes the values $\al=0.1$, $\al=0.2$ and $\al=0.3$ for the dotted, dashed and solid curves respectively and $n=5$. For the right graph we have taken $n=5$, $n=6$ and $n=7$ for the dotted, dashed and solid curves respectively and $\al=0.1$.}\label{Lt_graph}
\end{figure}
Using the relations (\ref{temp_x}), (\ref{v_2x}) and (\ref{press_x}) we can write it in evident form, but due to is complicated structure we do not show this evident form here, the dependence $L=L(T)$  is given graphically, namely the Figure [\ref{Lt_graph}] shows these dependences for various values of the coupling constant $\al$ and dimension $n$.  Both graphs on the Figure [\ref{Lt_graph}] demonstrate slight nonmonotonous dependence, which shows that for small temperatures  the latent heat increase as the temperature rises, whereas for higher temperatures the latent heat decreases when the temperature goes up. We also see that the latent heat increases in absolute values with increasing of the coupling parameter $\al$ and the  dimension $n$, but  general character of $L=L(T)$ dependence is the same. It should be also pointed out that similar behaviour of the latent heat  takes place for Einstein-Maxwell-dilaton black hole \cite{Stetsko_EPJC19}.

\section{Critical exponents}
In this section we examine the thermodynamic behaviour of the system in the vicinity of the critical point. The method we follow here was applied to black holes for the first time in the papers \cite{Kubiznak_JHEP12}, but it is well-known in the Condensed Matter Physics for a longer period of time \cite{Goldenfeld_92}.

Here we are going to obtain critical exponents which characterize the phase transition at the critical point. Firstly, to derive the critical exponent $\bar{\al}$ the entropy (\ref{entropy}) should be rewritten as a function of the temperature $T$ and thermodynamic volume $V$ (\ref{td_vol}):
\begin{equation}
S(T,V)=\frac{\omega_{n-1}}{4}b^{(n-1)\gamma}\left[\frac{(n-\alpha^2)}{\omega_{n-1}(1+\alpha^2)}b^{-(n-1)\gamma}V\right]^{(n-1)(1-\gamma)/((n-1)(1-\gamma)+1)}.
\end{equation}
Since there is no explicit temperature dependence, the heat capacity under the condition of the constant (fixed) volume is equal to zero ($C_V=0$) and as a consequence the critical exponent $\bar{\al}=0$. Here we also remark that for the mentioned above critical exponent we use the notation $\bar{\al}$ in order not to get confused with our dilaton coupling parameter which is denoted as $\al$.

The other critical exponents can be derived if one rewrites the equation of state (\ref{eq_of_st_2}) using the so called reduced variables:
\begin{equation}
p=\frac{P}{P_c}, \quad \tau=\frac{T}{T_c}, \quad \nu=\frac{v}{v_c},
\end{equation} 
and here $P_c$, $T_c$, $v_c$ are corresponding critical values defined by the relations (\ref{p_c}), (\ref{t_c}) and (\ref{v_c}) respectively. As a result, the equation of state takes the form:
\begin{equation}\label{red_eos}
p=\frac{4(2-\al^2)}{3(1-\al^2)^2}\frac{\tau}{\nu}-\frac{(1+\al^2)(2-\al^2)}{(1-\al^2)}\nu^{2(\gamma-1)}\left(1-\frac{\nu^{2(2\gamma-1)}}{3(2-\al^2)}\right).
\end{equation} 
Following the mentioned above procedure the written above equation can be cast in the form:
\begin{equation}\label{simpl_red_eos}
p=\frac{1}{\rho_c}\frac{\tau}{\nu}+h(\nu),
\end{equation}
where $\rho_c$ is the critical ratio given by the relation (\ref{cr_rat}) and $h(\nu)$ is the function which corresponds to the second term in the relation (\ref{red_eos}), thus the procedure is applicable for arbitrary form of the function $h(\nu)$. In the vicinity of the critical point we assume that $\tau=1+t$, $\nu=(1+\omega)^{\frac{1}{z}}$, where $t<0$ and $\omega<0$ and $z>0$. According to definition of the reduced variables at the critical point all of them are equal to one $p=\tau=\nu=1$, thus the new parameters $t$ and $\omega$ are equal to zero ($t=\omega=0$). Using these new variables $t$ and $\omega$ and expanding the right hand side of the relation (\ref{simpl_red_eos}) we write:
\begin{equation}\label{eq_exp}
p=1+At-Bt\omega-C\omega^3+{\cal O}(t\omega^2,\omega^4)
\end{equation}
and here 
\begin{equation}
A=\frac{1}{\rho_c}, \quad B=\frac{1}{z\rho_c}, \quad C=\frac{1}{z^3}\left(\frac{1}{\rho_c}-\frac{1}{6}h^{(3)}(1)\right),
\end{equation}
where $z=(n+\al^2)/(1+\al^2)$ and the parameter $C$ is also supposed to be positive. Differentiating the written above equation with respect to the parameter $\omega$ we arrive at the relation:
\begin{equation}\label{max_law}
{\rm d}P=-P_c\left(Bt+3C\omega^2\right){\rm d}\omega.
\end{equation} 
Taking into account the Maxwell's area law and denoting $\omega_l$ and $\omega_s$ the ``volumes'' of different phases (the large and small black holes) we can write:
\begin{equation}
p=1+At-Bt\omega_l-C\omega^3_l=1+At-Bt\omega_s-C\omega^3_s,
\end{equation} 
which has a nontrivial solution of the form:
\begin{equation}
\omega_s=-\omega_l=\sqrt{-\frac{Bt}{C}}.
\end{equation}
It allows us to obtain the following behaviour for the order parameter $\eta$ near the critical point and derive corresponding critical exponent $\bar{\beta}$:
\begin{equation}
\eta=V_c\left(\omega_l-\omega_s\right)=2V_c\omega_l\simeq(-t)^{1/2} \quad \Rightarrow \quad \bar{\beta}=\frac{1}{2}.
\end{equation}
Investigating the behaviour of the isothermal compressibility (\ref{iso_comp}) near the critical point we can find the respective critical exponent $\bar{\gamma}$, namely we can write:
\begin{equation}
\kappa_T=-\frac{1}{V}\left(\frac{\partial V}{\partial P}\right)_T\sim\frac{1}{P_cBt} , \quad \Rightarrow \quad \bar{\gamma}=1.
\end{equation}
Finally, using the equation of state (\ref{eq_exp}) and considering the critical isotherm ($t=0$) we obtain:
\begin{equation}
p-1=-C\omega^3, \quad \Rightarrow \quad \bar{\delta}=3.
\end{equation}
It should be pointed out that the critical exponents we have obtained here completely coincide with the corresponding values derived for Einstein-Maxwell-dilaton black hole \cite{Stetsko_EPJC19, Mo_PRD16,Dayyani_EPJC18} or even RN black hole \cite{Kubiznak_JHEP12}. 

\section{Conclusions}
In this work we have derived a static spherically symmetric black hole's solution in Einstein-Yang-Mills-dilaton theory. We take into account dilaton potential which is chosen in the so called Liouville form. It has been shown that the parameters of this potential are chosen to satisfy the field equations (\ref{einstein})-(\ref{scal_eq}). It should be also pointed out here that the Liouville-type potential allows not only to fulfill the field equations, but also to introduce an effective cosmological constant similarly as it was performed in Einstein-Maxwell-dilaton theory \cite{Sheykhi_PRD07,Stetsko_EPJC19}. As far as we know the Liouville-type potential was not used so often here as in case of Einstein-Maxwell-dilaton theory, namely it was utilized in the work \cite{Radu_CQG05_1}. The Einstein-Yang-Mills-dilaton theory with $SO(n)$ gauge group was examined \cite{Mazhari_GRG10}, but there Bertotti-Robinson-type solution was obtained which allowed the authors not to introduce dilaton potential $V(\Phi)$. The Yang-Mills potential was defined by virtue of the so-called magnetic Wu-Yang ansatz (\ref{gauge_pot}) \cite{Yasskin_PRD75,Mazhari_PRD07,Bostani_MPLA10}, which allowed us immediately satisfy the equations (\ref{YM_eq}). The obtained solution (\ref{metric_W}) is not very simple, but some general features stem immediately from its evident form. In particular it has the only simple root which is nothing else but the horizon point of the black holes, it is singular at the origin ($r\rightarrow 0$) and the behaviour of the metric function $W(r)$ at the origin is mainly defined by the Schwarzschild term which might be modified a bit by the Yang-Mills term. Finally, at the infinity the metric is not in strict sense anti-de Sitterian, but certainly, it is not asymptotically flat. To investigate the character of singular points that appear for our metric we have also calculated the Kretschmann scalar (\ref{Kr_scal}). It might be shown easily that the horizon point of the black hole is the point of ordinary coordinate singularity which might be removed by some new sort of coordinates as it is usually done for the most known types of black holes. The only point of physical singularity is the origin where the Kretschmann scalar has singular behaviour and which cannot be removed by any sort of new coordinate system. At the infinity the Kretschmann scalar was shown to go to zero while recovering well known result for AdS-case in the limit $\al\rightarrow 0$.

We have also studied thermodynamics of the black hole. Firstly, a standard approach has been utilized, namely the temperature (\ref{temp}), the entropy (\ref{entropy}) and the first law (\ref{first_law}) have been obtained. We note that the entropy equals to the quarter of the horizon area, similarly as we have for other types of black holes in the framework of standard General Relativity. The first law (\ref{first_law}) is completely identical to the corresponding relation for the standard Schwarzschild black hole and it is expectable since all the other parameters such as $\L$ or $q$ are supposed to be fixed. The temperature we have derived (\ref{temp}) shows nonmonotonous behaviour (what has been demonstrated clearly on the Figure [\ref{temp_graph}]) and this fact tells us about some unstability domains and also about possible criticality. We have also calculated the heat capacity (\ref{heat_capac}). The Figure [\ref{heatcap_gr}] shows that we have two discontinuity points which separate thermodynamically stable and unstable domains. We have also shown that increase of the cosmological constant $\L$ in absolute value gives rise firstly to merging of the discontinuities with their following transformation into a maximum point and if the module of $\L$ continues its rise the peak of this maximum finally disappears. In general this character of behaviour of the heat capacity $C_{\L,q}$ is identical to the situation we have for Einstein-Maxwell-dilaton black hole \cite{Stetsko_EPJC19}.

Using the extended phase technique we have also studied some aspects of thermodynamics. As it is usually performed \cite{Stetsko_EPJC19,Kubiznak_JHEP12,Kubiznak_CQG17,Dayyani_EPJC18} we introduce the thermodynamic pressure which is related to the cosmological constant $\L$. Since in our case we have additional parameter $q$ which as we have assumed gives the nonabelian charge and this value together with its conjugate are also taken into account in the extended phase space. Having used these additional thermodynamic values we have obtained the extended first law (\ref{ext_first_law}) and it allowed us to write the Smarr relation (\ref{smarr_gen}). We have also derived the equation of state (\ref{eq_of_state}). It is shown that the equation of state has an inflection point similarly as it takes place for Einstein-Maxwell-dilaton black holes \cite{Stetsko_EPJC19,Dayyani_EPJC18}, and corresponding critical values $v_c$, $T_c$, $P_c$ as well as their critical ratio (\ref{cr_rat}) have been calculated. What should be stressed here is the fact that the critical ratio depends on the coupling parameter $\al$ only and if $\al=0$ we recover the critical ratio for Van der Waals gas. In comparison the critical ratio for Einstein-Yang-Mills black hole \cite{Stetsko_EPJC19} depends on the dimension of space $n$ as well as on the parameter $\al$. We have calculated the Gibbs free energy (\ref{Gibbs_pot}) and we have shown that it has so-called swallow-tail behaviour for the temperatures below the critical one. It is demonstrated that for the temperatures below but quite close to the critical one there is a phase-transition of the zeroth order which was also observed in case of Einstein-Maxwell-dilaton theory \cite{Dayyani_EPJC18,Stetsko_EPJC19}. We have also found coexistence curve (Figure [\ref{coex_curves}]) and numerically calculated the latent heat (Figure [\ref{Lt_graph}]), and we can conclude that in general there are many common features with Einstein-Maxwell-dilaton black holes again. Finally we have calculated critical exponents $\bar{\al}$, $\bar{\beta}$, $\bar{\gamma}$ and $\bar{\delta}$ and as a result they are completely the same as for the Einstein-Maxwell-dilaton black hole,  or even for the Reissner-Nordstrom black hole \cite{Kubiznak_JHEP12}, but this fact can be easily explained, since the equations of state in their reduced forms are very similar  for all these cases. To sum all the described facts up we can state that being completely different solution in comparison with the mentioned many times Einstein-Maxwell-dilaton black holes the thermodynamics for the both Einstein-Maxwell-dilaton and Einstein-Yang-Mills-dilaton cases shows a lot of similarities.

\section{Acknowledgments}
This work was partly supported by Project FF-83F (No. 0119U002203) from the Ministry of Education and Science of Ukraine.  

\end{document}